\documentclass[conference]{IEEEtran}

\usepackage{amsfonts}
\usepackage{times}
\usepackage{graphicx}
\usepackage{latexsym}
\usepackage{amsmath}
\usepackage{graphicx}

\usepackage{subfigure}

\IEEEoverridecommandlockouts

\begin{document}

\title{Iterative Decoding and Turbo Equalization: The Z-Crease Phenomenon\thanks{Li's work is supported by the US National Science Foundation under the Grants No. CCF-0928092, CMMI-0829888 and OCI-1122027.}}
\author{Jing Li (Tiffany) \ \ and \ \   Kai Xie \\ 
 Department of Electrical and Computer Engineering, 
Lehigh University, Bethlehem, PA, 18015, US\\
Emails: \{jingli, kax205\}@ece.lehigh.edu
\vspace{-0.6cm}
}

\maketitle

\begin{abstract}

Iterative probabilistic inference, popularly dubbed the soft-iterative paradigm, has found great use in a wide range of communication applications, including turbo decoding and turbo equalization. The classic approach of analyzing the iterative approach inevitably use the statistical and information-theoretical tools that bear ensemble-average flavors. 
 This paper consider the per-block error rate performance, and analyzes it using nonlinear dynamical theory. By modeling the iterative processor as a
nonlinear dynamical system, we report a universal ``Z-crease phenomenon:''  the zig-zag or up-and-down fluctuation -- rather than the monotonic decrease -- of the per-block errors,  as the number of iteration increases. 
 Using the turbo decoder as an example, we also report several interesting motion phenomenons which were not previously reported, and which appear to correspond well with the notion of ``pseudo codewords'' and ``stopping/trapping sets.''  
We further propose a heuristic stopping criterion to control Z-crease and identify the best iteration. Our stopping criterion is most useful for controlling the worst-case per-block errors, and helps to significantly reduce the average-iteration numbers.

\end{abstract}


\vspace{-0.3cm}
\section{Introduction}
\label{sec:introduction}

The discovery of turbo codes and the re-discovery of low-density parity-check (LDPC) codes have, over the night, closed the theory-practice gap of the Shannon capacity limit on additive white Gaussian noise (AWGN) channels. They have also revolutionized the coding research with a new paradigm of {\it iterative probabilistic inference}, commonly dubbed the {\it soft-iterative} paradigm.
 Since their success with turbo codes and LDPC codes, the soft-iterative paradigm has become a vital tool in widespread applications in communication and signal processing.  A complex communication system comprised of layers of functional blocks that were previously individually or sequentially tackled, can now use a ``soft-iterative'' treatment, close in spirit to that of the turbo or LDPC decoder, to achieve quality performance with manageable complexity. Celebrated applications include, for example, iterative demodulation and decoding, turbo equalization (also known as iterative decoding and equalization),
and multi-user detection, and iterative sensing and decision fusion for sensor networks. 

The significance and wide popularity of the soft-iterative algorithm has caused a considerable amount of study on its behavior, performance and convergence. Present models and methodologies for analyzing estimation and decoding methods may be roughly grouped into the following categories:

1). The most straight-forward way to evaluate the performance of an estimation/detection/decoding method, be it soft-iterative or otherwise, is through Monte Carlo simulations. The result is very accurate, but the simulation is usually lengthy, tedious, and not scaling well.  
Additionally, simulations do not shed much insight into why the performance is so and how the performance might be improved.

2). Classic analytical methods come from the perspectives of maximum likelihood (ML), maximum {\it a posteriori} (MAP) probability, or minimum square error (MSE). They inevitably assume that the subject method is optimal and always deciding on the candidate that has the largest probability, maximum likelihood ratio, or the minimal Hamming/Euclidean distance to what's been observed.  
They produce  useful performance bounds, but may present a non-negligible gap to the true performance of the practical, iterative estimator at hand.

3). Powerful iterative analytical methods, notably the {\it density evolution} (DE) \cite{bib:DE} and the {\it extrinsic information transfer} (EXIT) charts \cite{bib:EXIT}, were developed in the last decade. These methods faithfully capture the iterative trajectory of many estimators/decoders, and have unveiled several fundamental and intriguing properties of the system (e.g. the ``convergence property'' and the ``area property'') \cite{bib:EXIT}. However, several underlying assumptions thereof, including the  ergodicity assumption, the neighborhood independence assumption and the Gausianity assumption, make them suitable mostly for evaluating the {\it asymptotic} behavior (i.e. infinite block size). Many real systems have limited lengths of a few hundred to a few thousand (bits), and the accuracy and usefulness of these methods can become limited in such cases.

4). To tackle the hard problem of iterative analysis for short-length signal sequence, researchers have also developed several interesting concepts and ideas, including {\it pseudo codewords} \cite{bib:pseudo}, {\it stopping sets} and {\it trapping sets} \cite{bib:stopping} \cite{bib:trapping}. 
They boast some of the most accurate performance predictions at short lengths. 
The drawback, however, is that efficient and systematic ways to identify and quantify these metrics are not readily existent, 
and hence, one may have to rely on computer-aided search of some type, causing daunting complexity.

The majority of the existing methods, as summarized above, largely stem from a statistical and/or information theoretical root. They have significantly advanced the field, but are also confronted with challenges and limitations, as they try to use statistical metrics and tools that are based on ensemble averages (such as mean, variance, entropy, mutual information) to predict and control the iterative process of a large-dimension, highly-dynamical, and apparently-random signal sequence.

The notion that iterative probabilistic inference algorithms can be viewed as complex dynamical systems \cite{bib:Richardson first dynamical analysis for turbo}-\cite{bib:He dynamic analysis for Turbo product code} presents an interesting departure from the existing ensemble-average based methods, and brings up new ways of evaluating the soft-iterative algorithm on individual blocks. 
Generally speaking, the performance of a system should be assessed  
in the {\it average} sense, such as the bit error rate (BER) averaged 
over hundreds of thousands of blocks. At the same time, however, it also makes sense to evaluate the {\it per-block} performance, namely, the number
(or the percentage) of errors in individual blocks. Per-block error
rate reflects error bursts and/or the worst-case situation, and can be
of interest in several applications. For example, in multimedia
transmission, a modest number of bit blip errors in an image  may
cause only a minor quality degradation that is hard to perceive by
human eyes, whereas excessive errors will cause the image to be
badly distorted and unusable. In magnetic recording systems, a Reed
Solomon (RS) outer code  is generally employed after the channel-coded
partial-response (PR) channel, to clear up the residual errors left
by the  equalizer/decoder. It does not really matter that every block has errors (after the equalizer/decoder); as long as the per-block error rate is within the error correction capability of the RS wrap, zero-error is achievable for
the entire system.
Hence, the specific issue we investigate here is how the per-block performance improves with the number of iterations.

Common wisdom has it that more iterations can not hurt, i.e., a larger number of iterations may not necessarily lead to (worthy or noticeable) performance gains, but it cannot degrade the performance either.  This apparent truth, as verified by the numerous studies reported on the bit error rate (BER) simulations, density
evolution (DE) analysis, and extrinsic information transfer (EXIT)
charts, holds in the context of {\it average} performance.  
In terms of the per-block performance, however, our studies reveal that it is not only possible, but also quite likely, for an individual block to encounter an ``fluctuating'' decoding state, such that the number of errors in that block keeps bouncing up and down (rather than monotonically decreasing) with the iterations.
 What this phenomenon,
thereafter referred to as the {\it Z-crease} phenomenon, implies in
practice is that a larger number of iterations are not always beneficial, and that the right timing may play a more important role. 
If the decoder stops at an unlucky iteration, it may actually
generate far more errors than if it stopped several iterations
earlier. 

It is worth noting that Z-crease is not
special; it is actually a {\it universal} phenomenon that vastly exists
in iterative decoding and estimation systems. We have examined a variety of different
systems, including low-density parity-check (LDPC) codes with
message-passing decoding, turbo codes with turbo decoding, product
accumulate (PA) codes with iterative PA decoding \cite{bib:PAcodes}, and
convolutionally-coded inter-symbol interference (ISI) channel with turbo equalization. In all of
these systems, we have observed the Z-crease phenomenon. 



To study the per-block system behavior and reveal the Z-crease phenomenon, our approach is to treat the iterative estimation/decoding system a high-dimensional nonlinear dynamical system parameterized by  a set of parameters, to further transform it to a one-dimensional state space with a single parameter, and examine the time evolution of the states at specific signal-to-noise ratios (SNR). 
 In this paper, we take a popular turbo code as an example, and report here the observation of a wide range of phenomena characteristic to nonlinear dynamical systems, including several new phenomena not reported previously.  

\section{Modeling Turbo Decoder as Nonlinear Dynamical System}

 A communication or signal processing system generally consists of many processing blocks inter-connected in parallel, in serial, or in hybrid, each fulfilling a specific task. 
Since the solution space of an ``integrated'' process is the Kronecker product of all the constituent solution spaces, to launch an overall optimal solution usually induces prohibitive complexity. A more feasible solution is to apply iterative algorithms, which allow constituent sub-units to perform local process and to iterative exchange and refine processed ``messages'', thus achieving a solution considerably better than that from sequential processing with a manageable complexity.

Iterative algorithms are by nature probabilistic inference based, where the ``messages'' to be processed and communicated represent the reliability or confidence level of a digital decision, commonly formulated as {\it log-likelihood ratios (LLR)}; but they can also be modeled as (nonlinear) dynamical systems. 
To help model all the variants of iterative algorithms in a universal mathematical formulation, we have summarized some the properties assumed to be features of these algorithms: (i) An iterative algorithm is a dynamical system with a large number of dimensions, possibly depending on many parameters, and distances along trajectories increase (decrease) polynomially, sub-polynomially or exponentially. (ii) It is formed by two or more units interacting with each other; each unit responds to messages received from the others in a nonlinear manner. (iii) The system is in general hierarchical: a message may be treated in several different levels (units) before reaching the center of action. (iv) The system in its evolution may be adaptive, i.e. with memory. (v)  Local interactions may have global effect: they may produce considerable global change in the system over the time, e.g. the ``wave effect'' in the decoding of an irregular LDPC code.

\begin{figure}[h]
\centerline{
\includegraphics[width=2.2in]{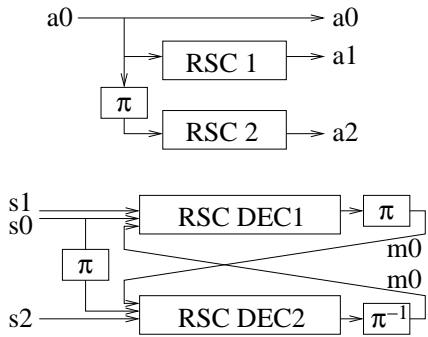}
\vspace{-0.2cm}
}
\label{figure:1turbo}
\caption{Turbo codes. Top: turbo encoder; Bottom: turbo decoder.}
\end{figure}

We start by evaluating the turbo decoder, which is useful in its own right\footnote{Turbo codes are in a number of standards, including the 3rd Generation Cellular Networks and Digital Video Broadcasting (DVB).}, and whose information theoretical analysis has reached a good level of maturity. A typical rate-1/3 turbo code, depicted in Fig. \ref{figure:1turbo}, is formed of two constituent recursive systematic convolutional (RSC) codes, concatenated in parallel through a pseudo-random interleaver. It encodes a block of $k$ binary bits ${\bf a}_0$ into a codeword  of $3k$ binary bits, $[{\bf a}_0, {\bf a}_1,{\bf a}_2]$. The decoder operates much like the turbo engine in an automobile, in which two sub-decoders perform soft-in soft-out decoding, and iteratively exchange and refine LLR messages ${\bf m}_0$ corresponding to ${\bf a}_0$. Let $[{\bf z}_0, {\bf z}_1, {\bf z}_2]$ be the noise, induced by the physical channel or transmitter/receiver circuitry, and let ${\bf s}_i={\bf a}_i+{\bf z}_i$ be the noisy observation available at the decoder. Exploiting the geometric uniformity of the codeword space of a turbo code (or any practical ECC), we can model the turbo decoder as a discrete-time dynamical system in constant evolution: \vspace{-0.1cm}
\begin{align}
{\bf m}_0^{<n+1>}&={f}_1({\bf s}_0, {\bf s}_2; {\bf m}_0^{<n>} )=
{f}_1({\bf z}_0, {\bf z}_2; {\bf m}_0^{<n>} ), \ \ \ \label{eqn:1}\\
{\bf m}_0^{<n+2>}&={f}_2({\bf s}_0,{\bf s}_1; {\bf m}_0^{<n+1>})=
{f}_2({\bf z}_0, {\bf z}_1; {\bf m}_0^{<n+1>} ), \label{eqn:2}
\end{align}
where the superscript $n$ denotes the number of half iterations, ${\bf z}_0$, ${\bf z}_1$ and ${\bf z}_2$ are the parameters of the dynamical system,  and ${f}_1$ and ${f}_2$ are nonlinear functions 
describing the constituent RSC sub-decoders, reflecting in general an implementation of the Bahl-Cocke-Jelinek-Raviv (BCJR) decoding algorithm, the soft Viterbi algorithm (SOVA), or their variations. 

When $k$ takes on a value of a few thousand or larger, as in a practical scenario, this $k$-dimensional $3k$-parametrized nonlinear dynamical system becomes too complex to characterize or visualize.  
To make the problem tractable, we propose to ``project'' these dimensions to one or a few ``critical'' ones. Borrowing insight developed from conventional decoder analysis and after performing a careful evaluation, we propose to project the $3k$ parameters into a single parameter, the (approximated) signal-to-noise ratio (SNR), $\gamma=\frac{3k}{2R}/|| \,[{\bf z}_0,{\bf z}_1,{\bf z}_2]\, ||^2$, (where $R=1/3$ is the code rate), and to project the $k$ dimensions of the state space into one dimension, the  mean magnitude of LLRs, $x_0^{<n>}={\bf E}[|m_i|]$ where $m_i\in {\bf m}_0^{<n>}$. The nonlinear dynamical system remains in constant evolution in the $k$-dimensional space with $3k$ parameters (as a real turbo decoder  does), but characterizing the system using reduced dimensions drastically simplifies the analysis, enabling a better visualization and understanding of the further behavior. 

\vspace{-0.2cm}
\vspace{-0.3cm}    

\section{Analyzing Turbo Decoder Using Nonlinear Dynamical Theory}

\subsection{Phase Trajectories and Nonlinear Dynamical Behavior}

Consider noise samples represented by
$[{\bf z}_0, {\bf z}_1, {\bf z}_2] = [z_1, z_2, . . . , z_{3k}]$. Different vectors of noise samples are said to have the same noise realization, if they have the same fixed ratios between consecutive
sample values, $z_1/z_2$, $z_2/z_3$, $\cdots$, $z_{3k-1}/z_{3k}$. 
Thus for a given noise realization, the noise vector ${\bf z}_{0,1,2}$ is completely determined by the (approximated) SNR $\gamma$.
In general, $k$ should be chosen sufficiently large to make $\gamma$ a close approximation of the true channel SNR.

\begin{figure}[htbf]
\centerline{
\includegraphics[width=1.8in]{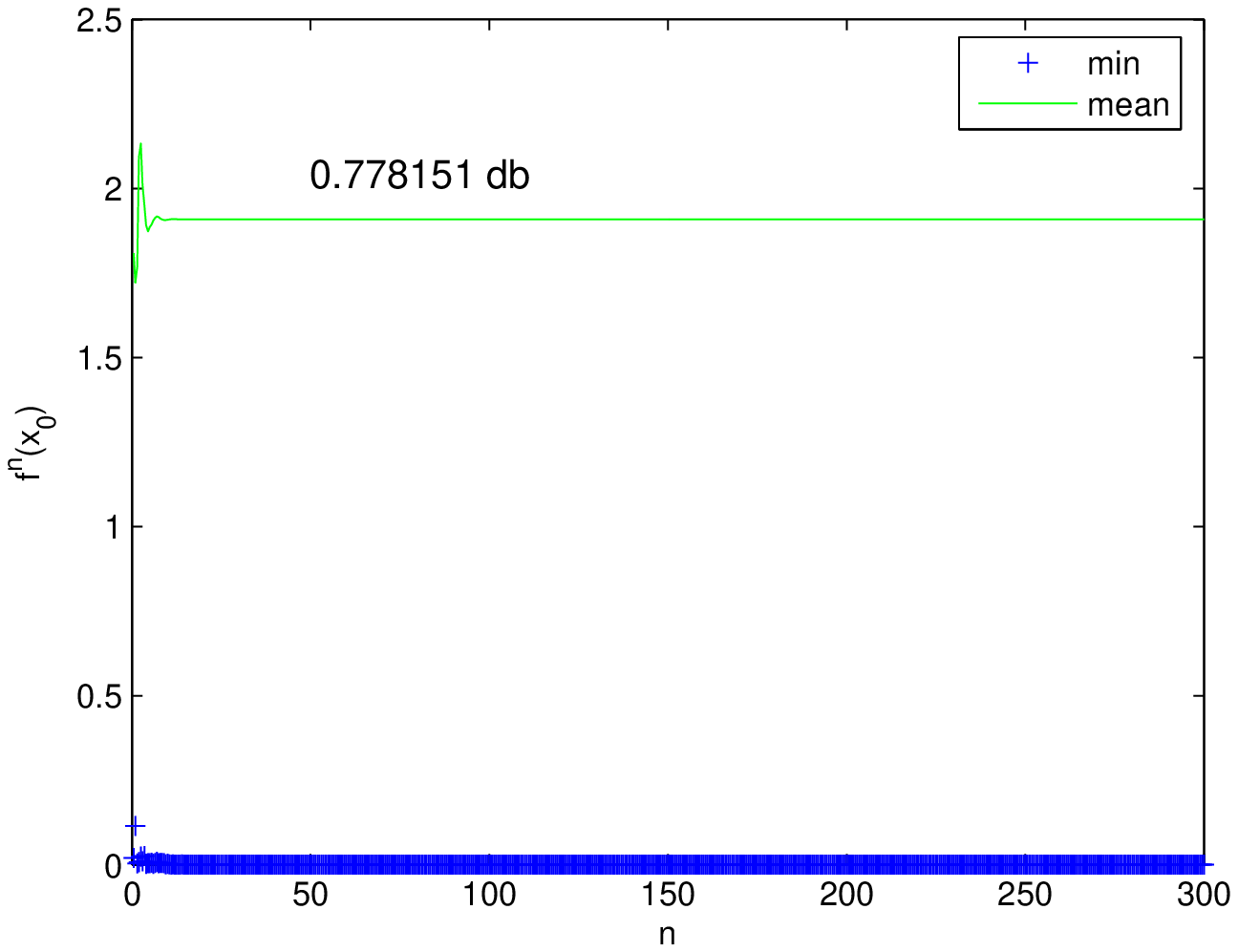}
\includegraphics[width=1.8in]{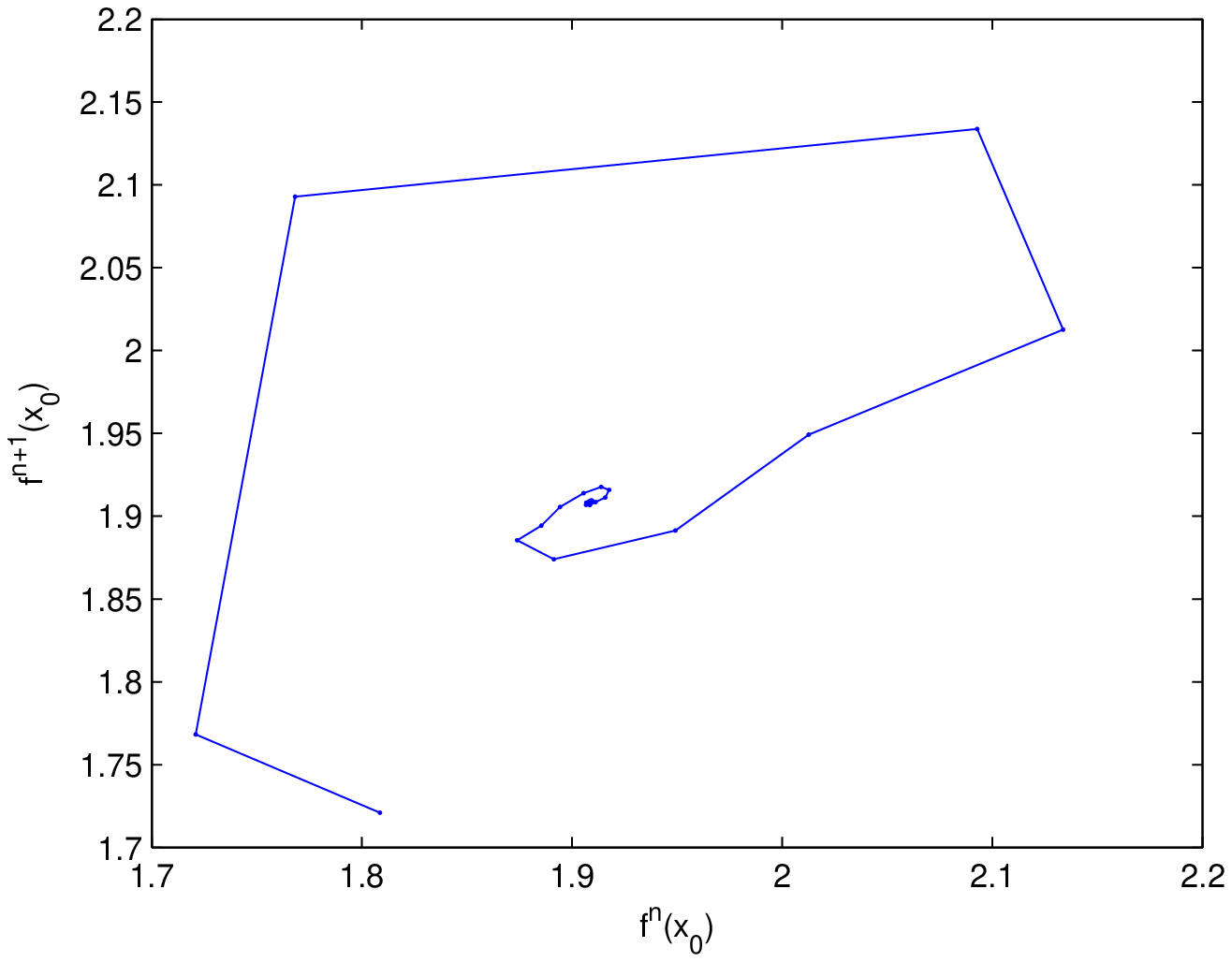}
}
\vspace{-0.2cm}
\caption{ $\gamma=0.778151$ db, indecisive fixed point. (Left: wave picture; Right: trajectory picture.)}
\label{fig:2a}
\end{figure}

\begin{figure}[htbf]
\centerline{
\includegraphics[width=1.8in]{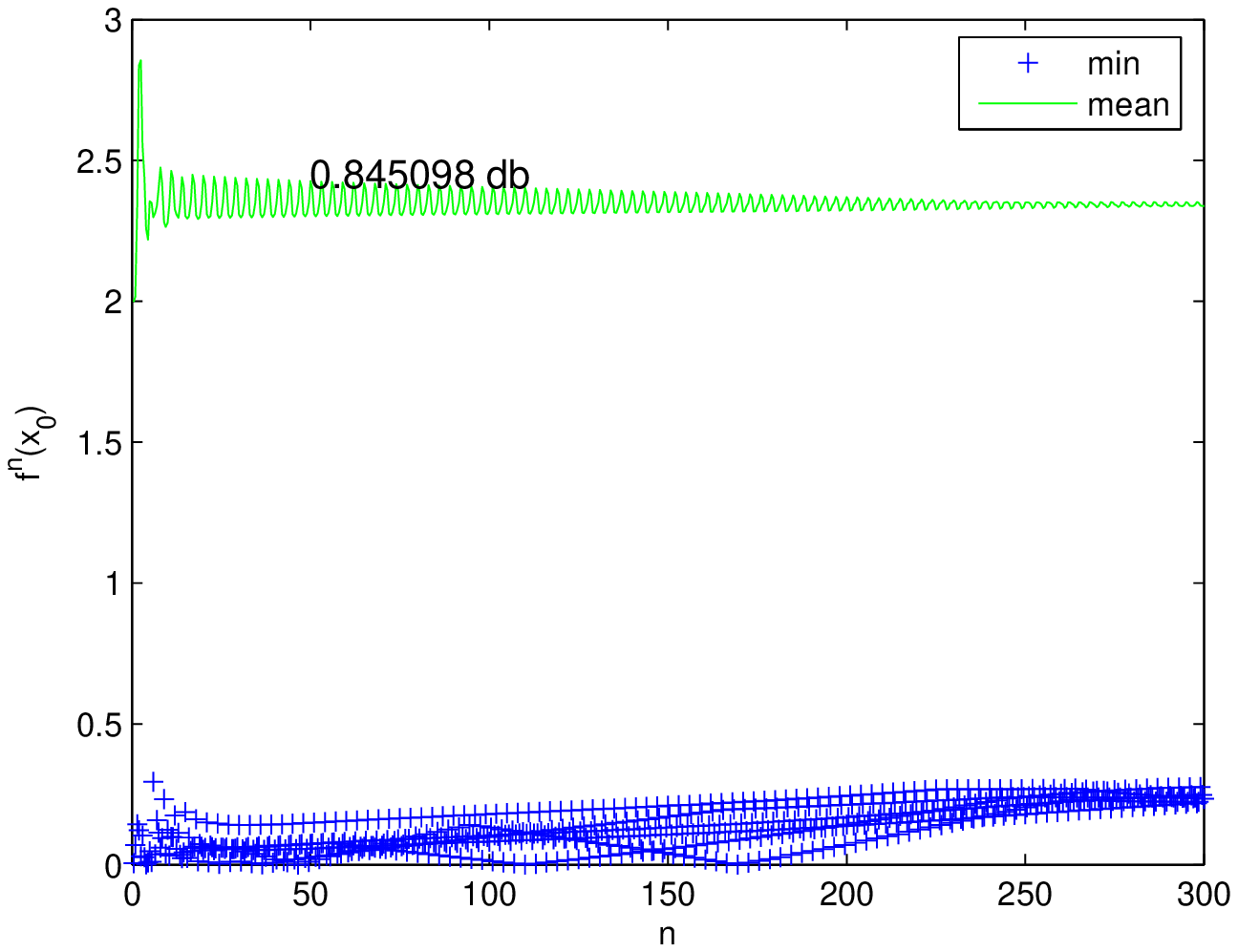}
\includegraphics[width=1.8in]{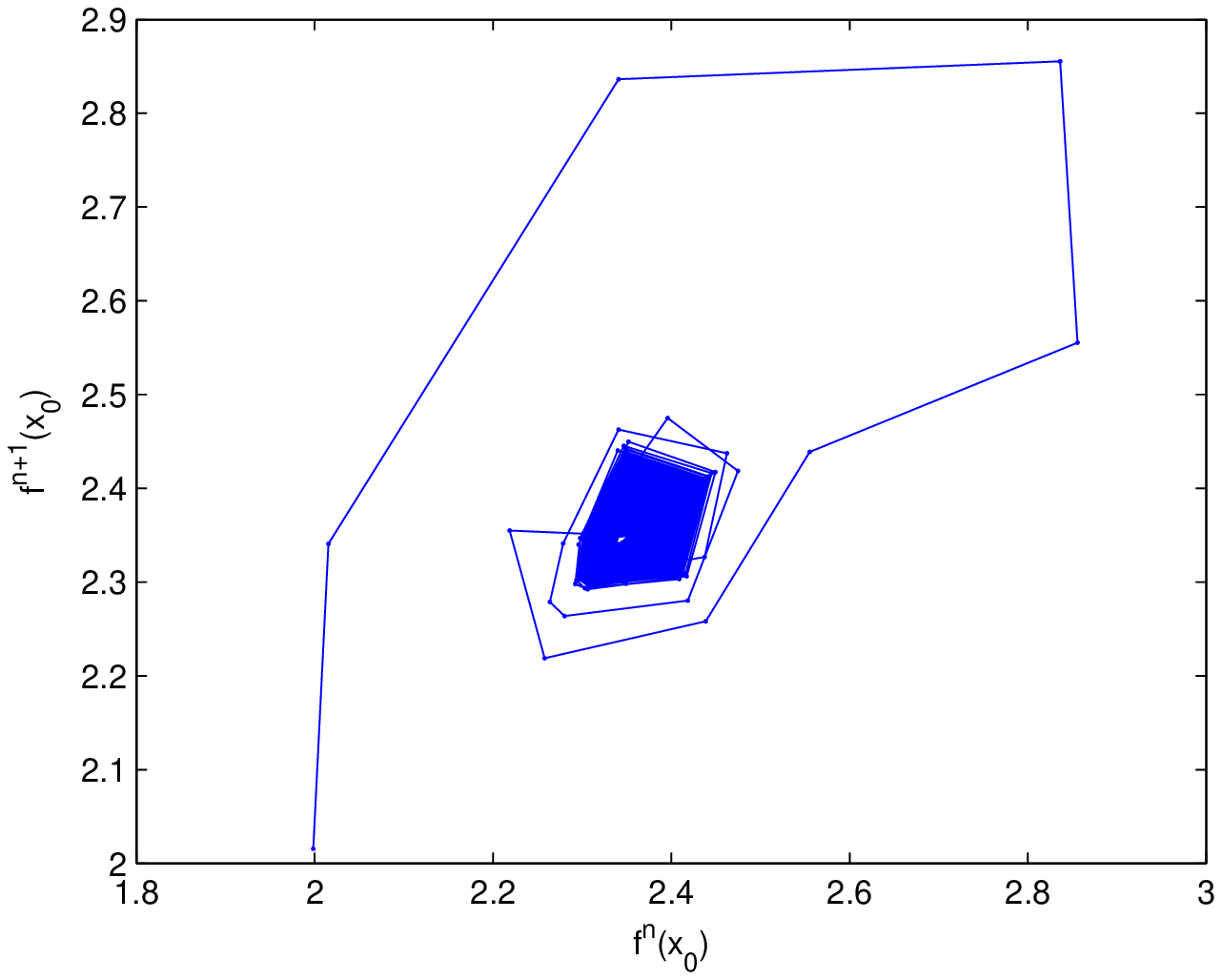}
}
\vspace{-0.4cm}
\label{fig:2b}
\caption{ $\gamma=0.845098$ db, fixed point breaking down.}
\vspace{0.2cm}
\centerline{
\includegraphics[width=1.8in]{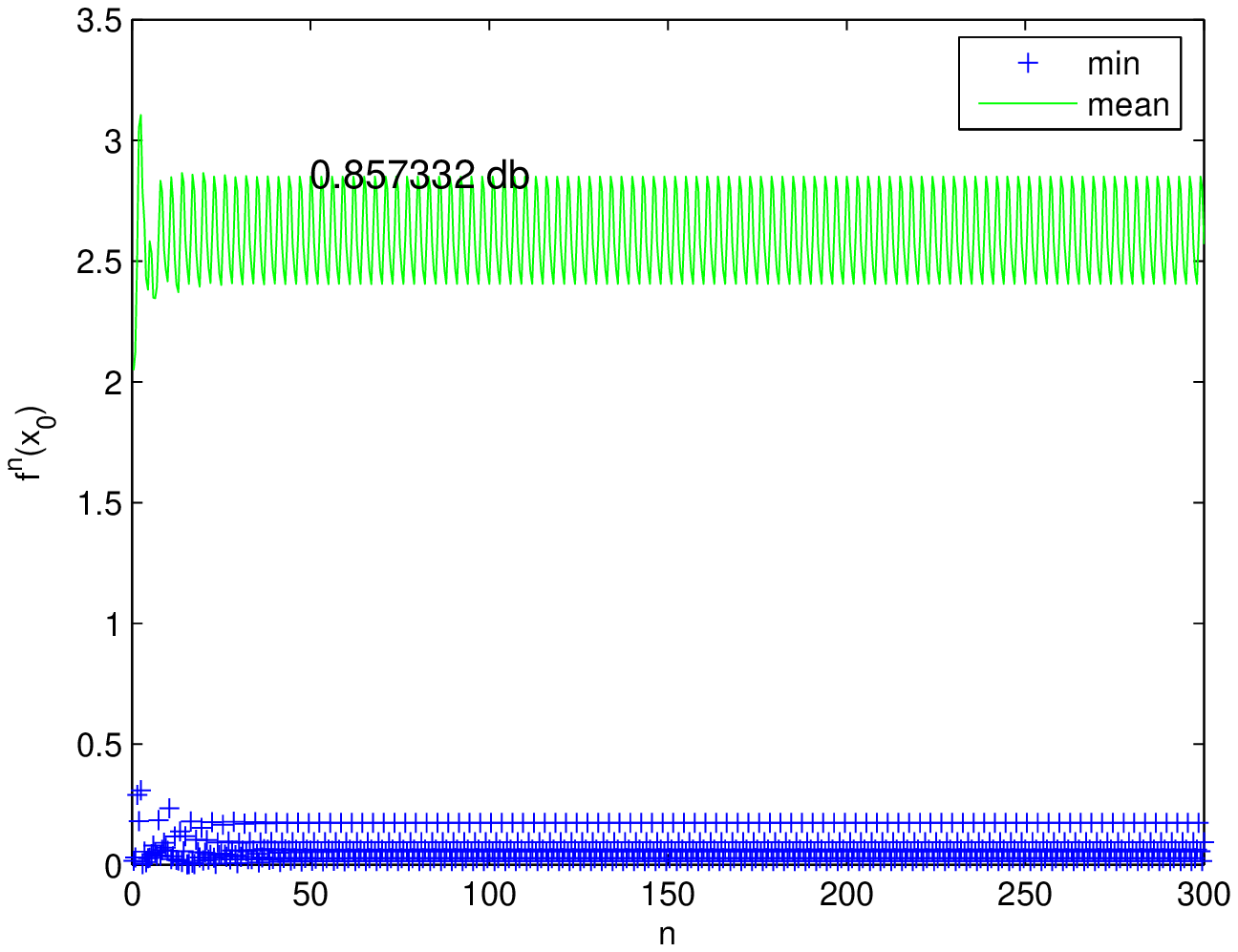}
\includegraphics[width=1.8in]{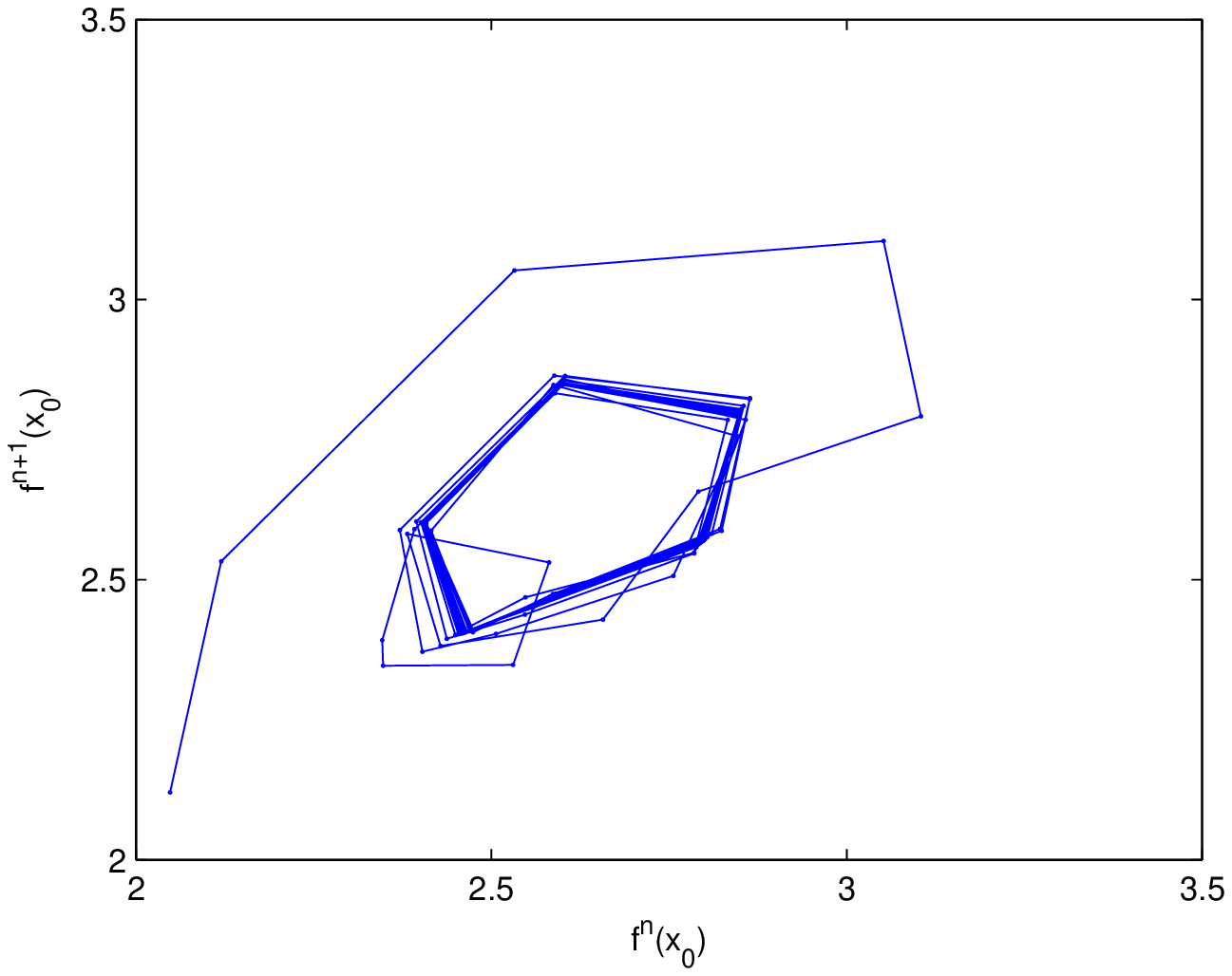}
}
\centerline{
\includegraphics[width=1.8in]{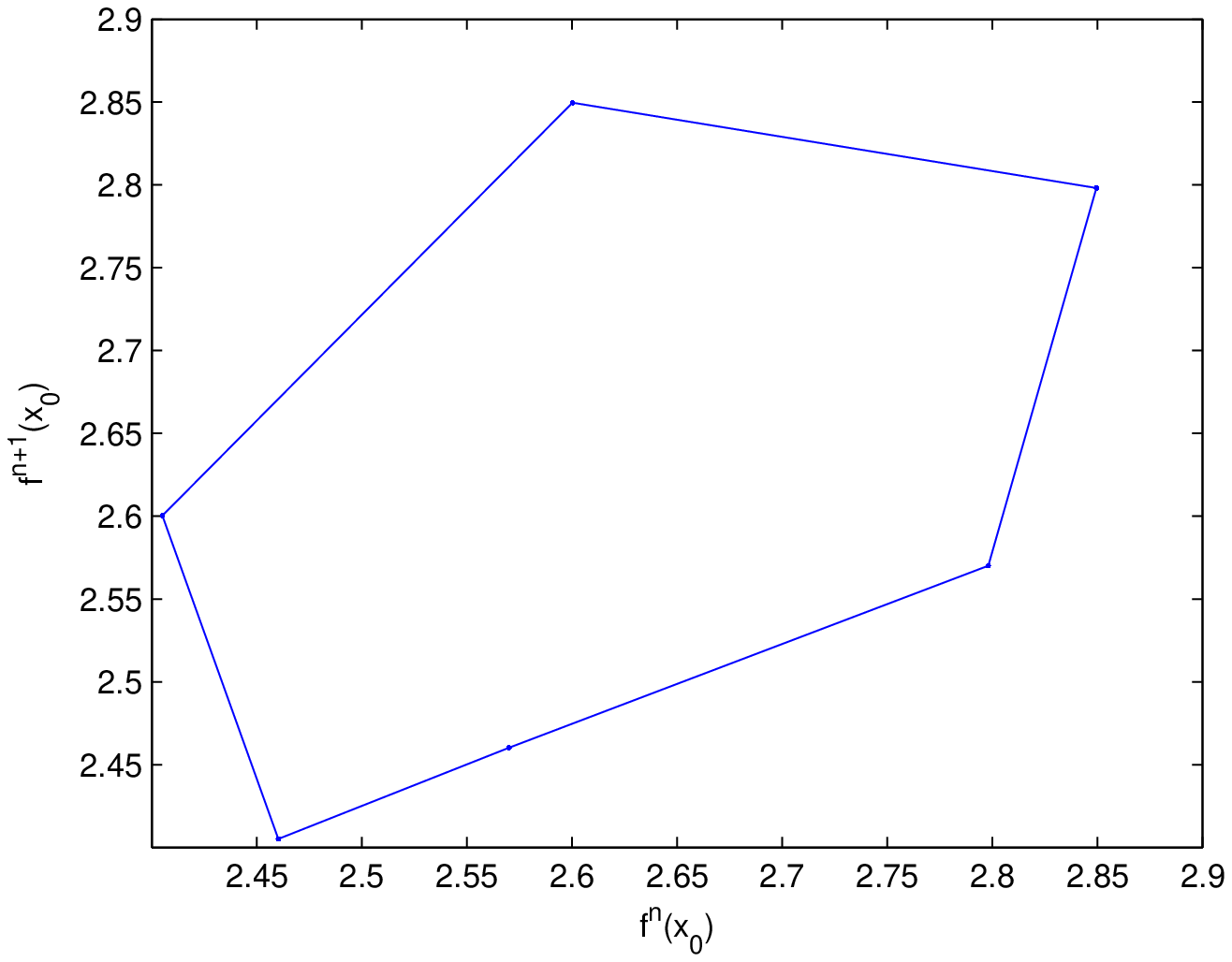}\hspace{1.8in}
}
\vspace{-0.4cm}
\caption{$\gamma=0.857332$ db, periodical fixed point.
(Top Left: wave picture; Top Right: trajectory picture; Bottom Left: zoomed in trajectory picture.)}
\label{fig:2c}
\vspace{0.2cm}
\centerline{
\includegraphics[width=1.8in]{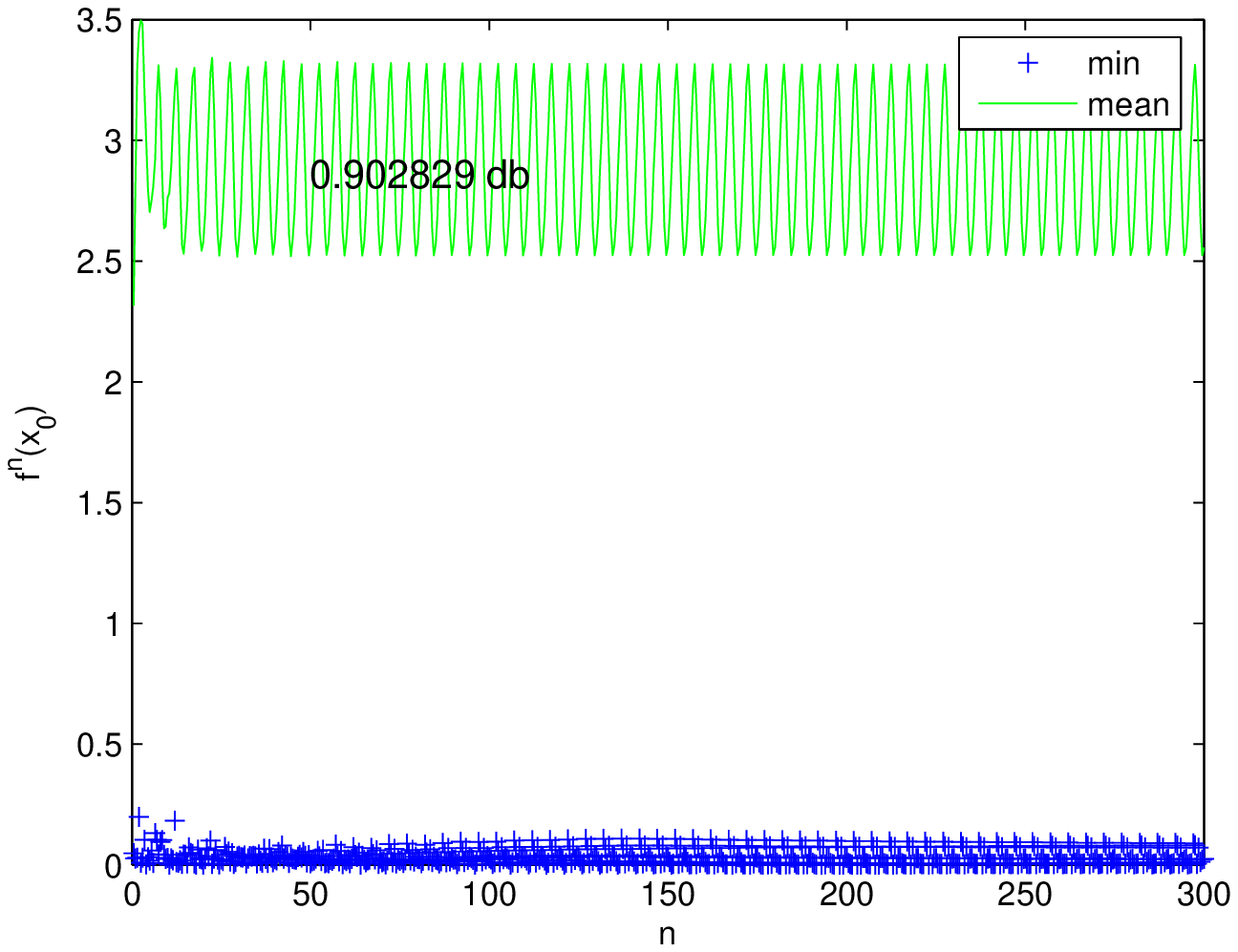}
\includegraphics[width=1.8in]{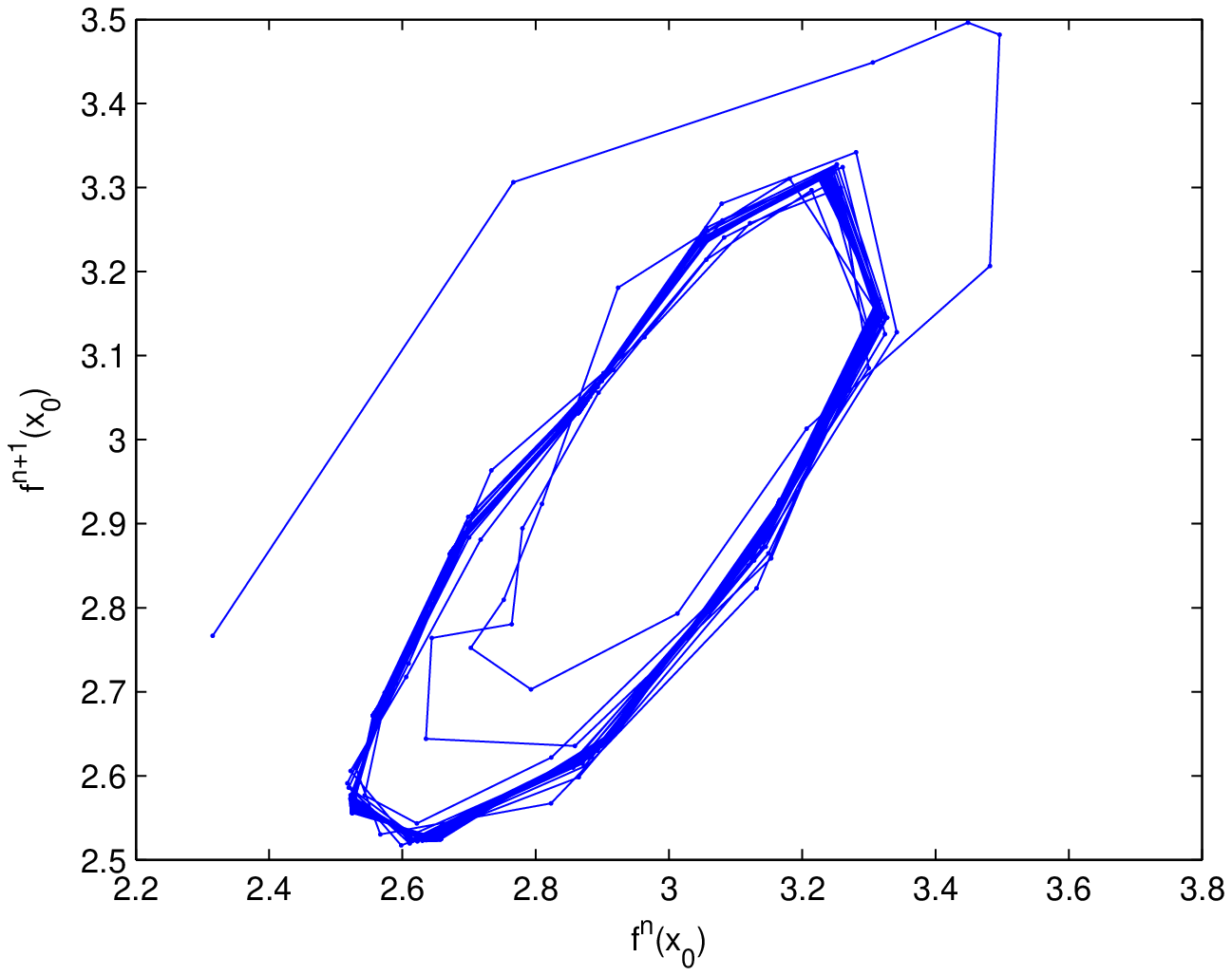}
}
\centerline{
\includegraphics[width=1.8in]{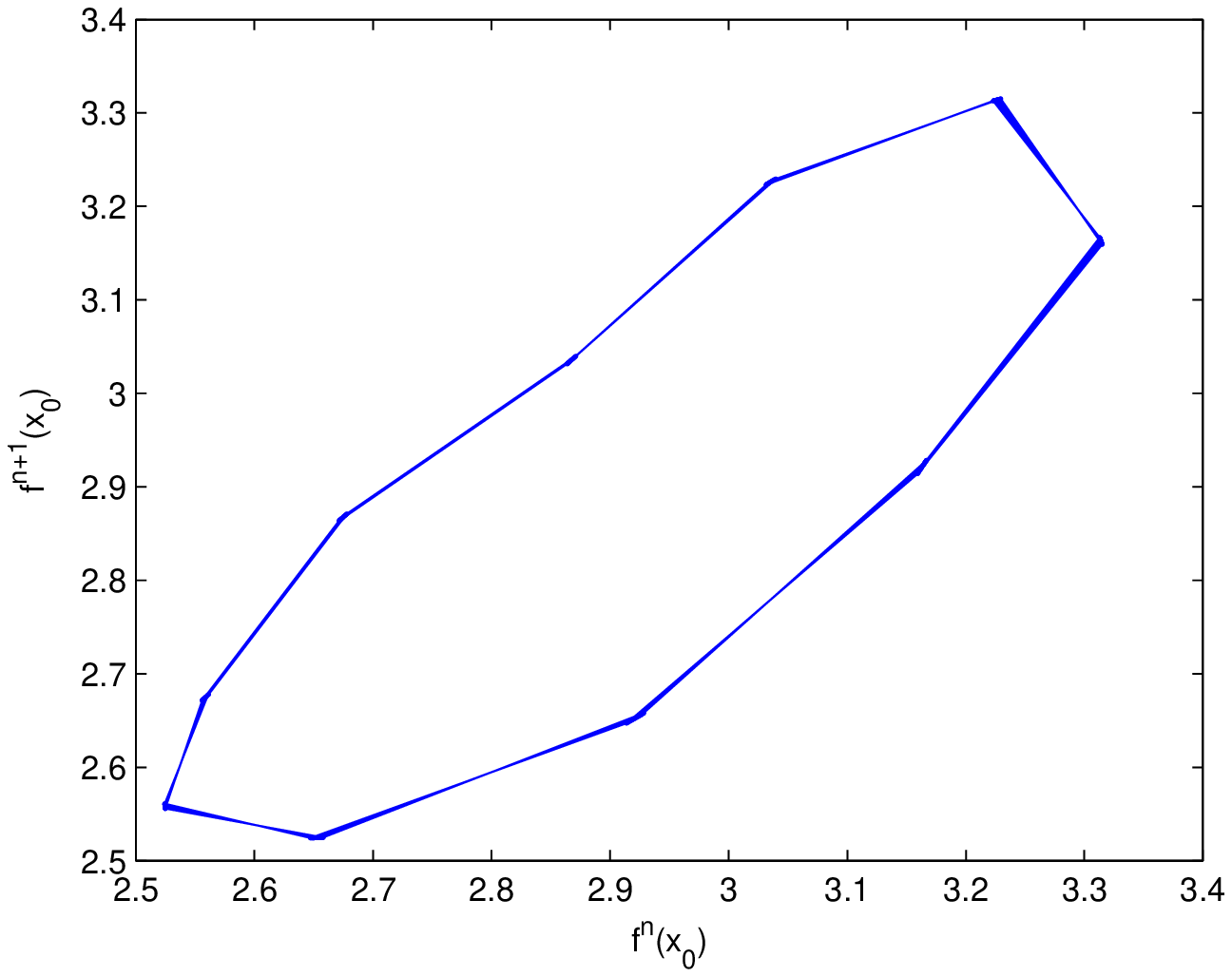}\hspace{1.8in}
}
\vspace{-0.4cm}
\caption{$\gamma=0.902829$ db, periodical fixed point losing stability.}
\label{fig:2d}
\vspace{-0.4cm} 
\end{figure}

\begin{figure}[htbf]
\centerline{
\includegraphics[width=1.8in]{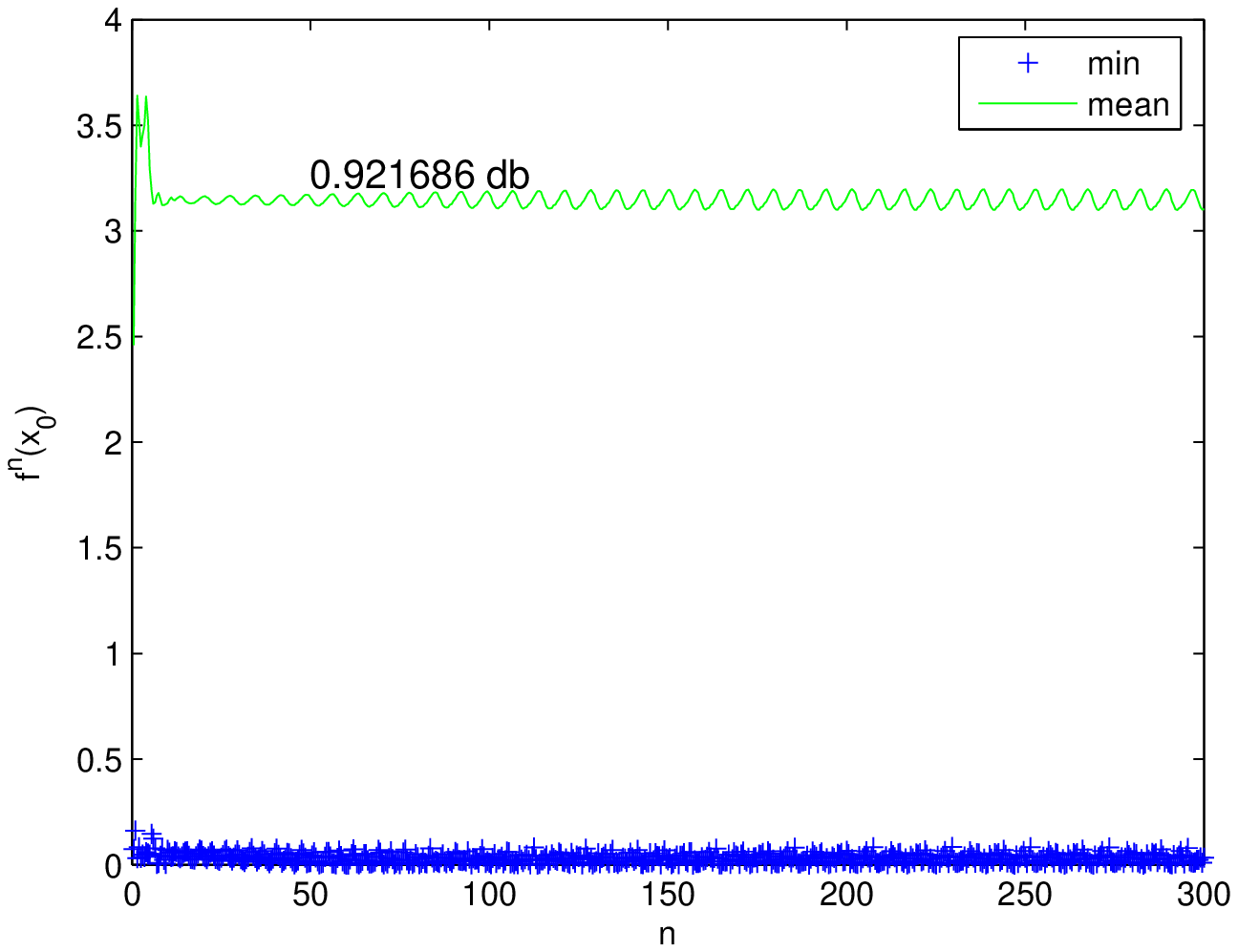}
\includegraphics[width=1.8in]{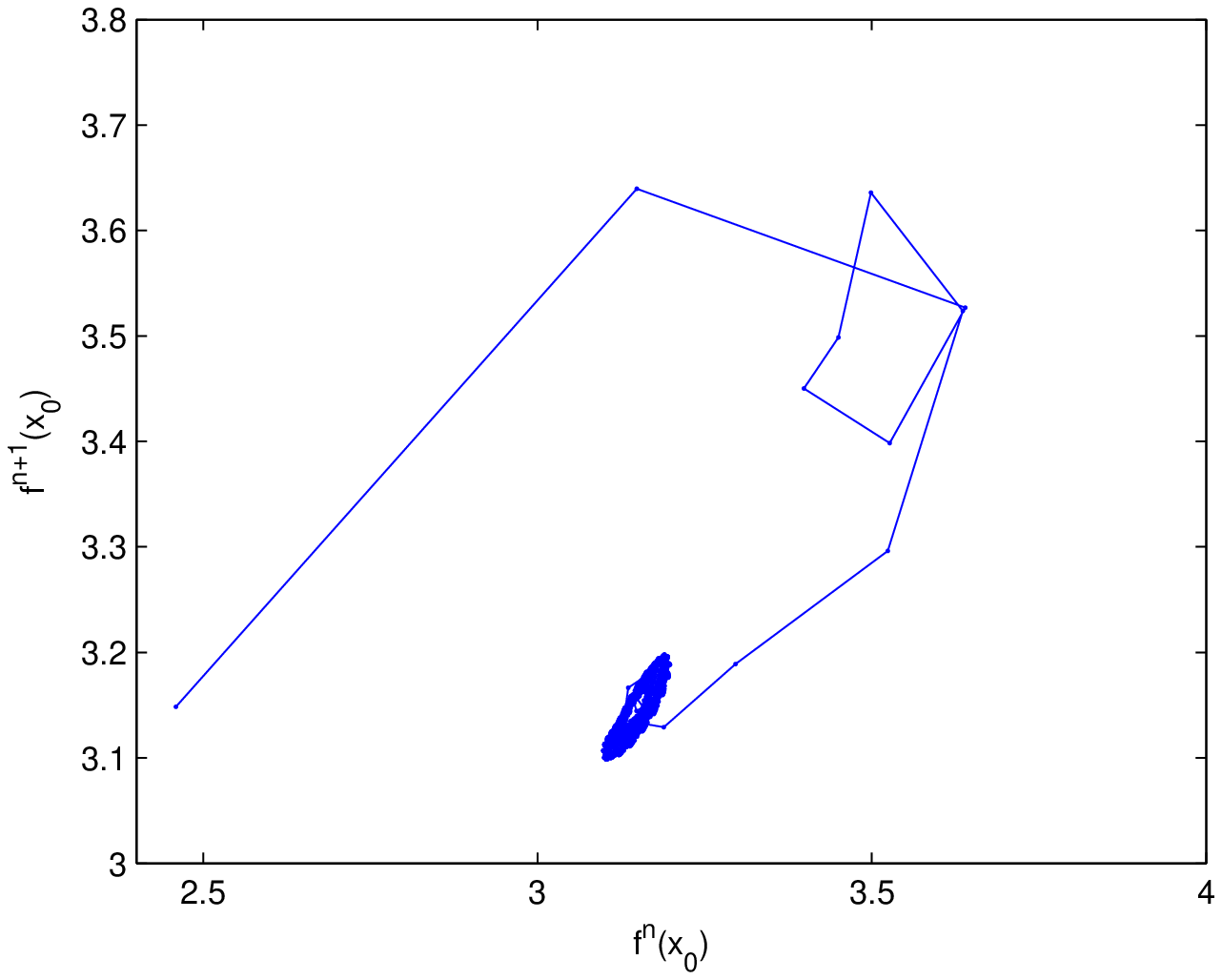}
}
\centerline{
\includegraphics[width=1.8in]{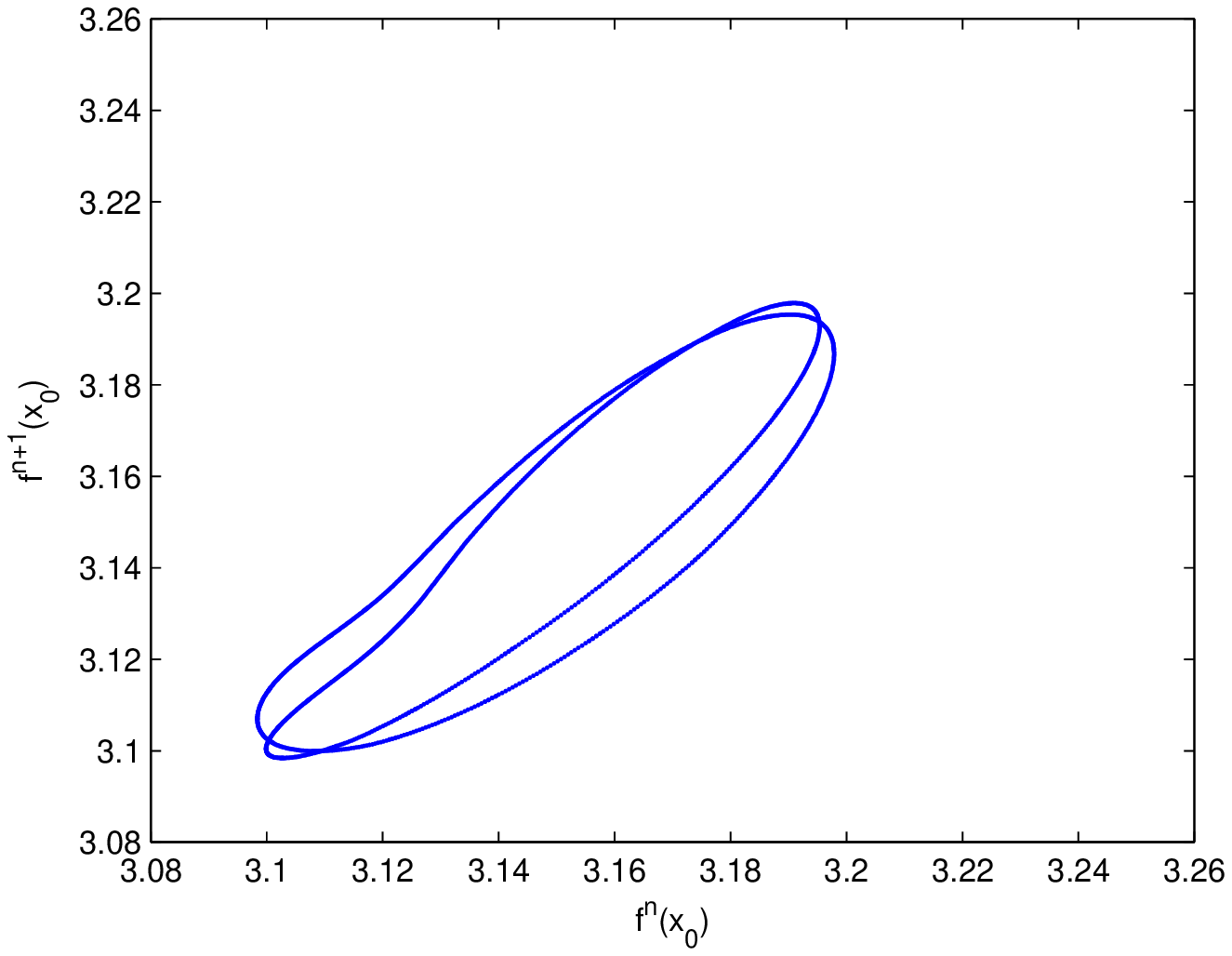}\\ \hspace{1.8in}
}
\vspace{-0.4cm}
\caption{$\gamma=0.921682$ db, limited-cycle tractor.}
\label{fig:2e}
\vspace{0.3cm}
\centerline{
\includegraphics[width=1.8in]{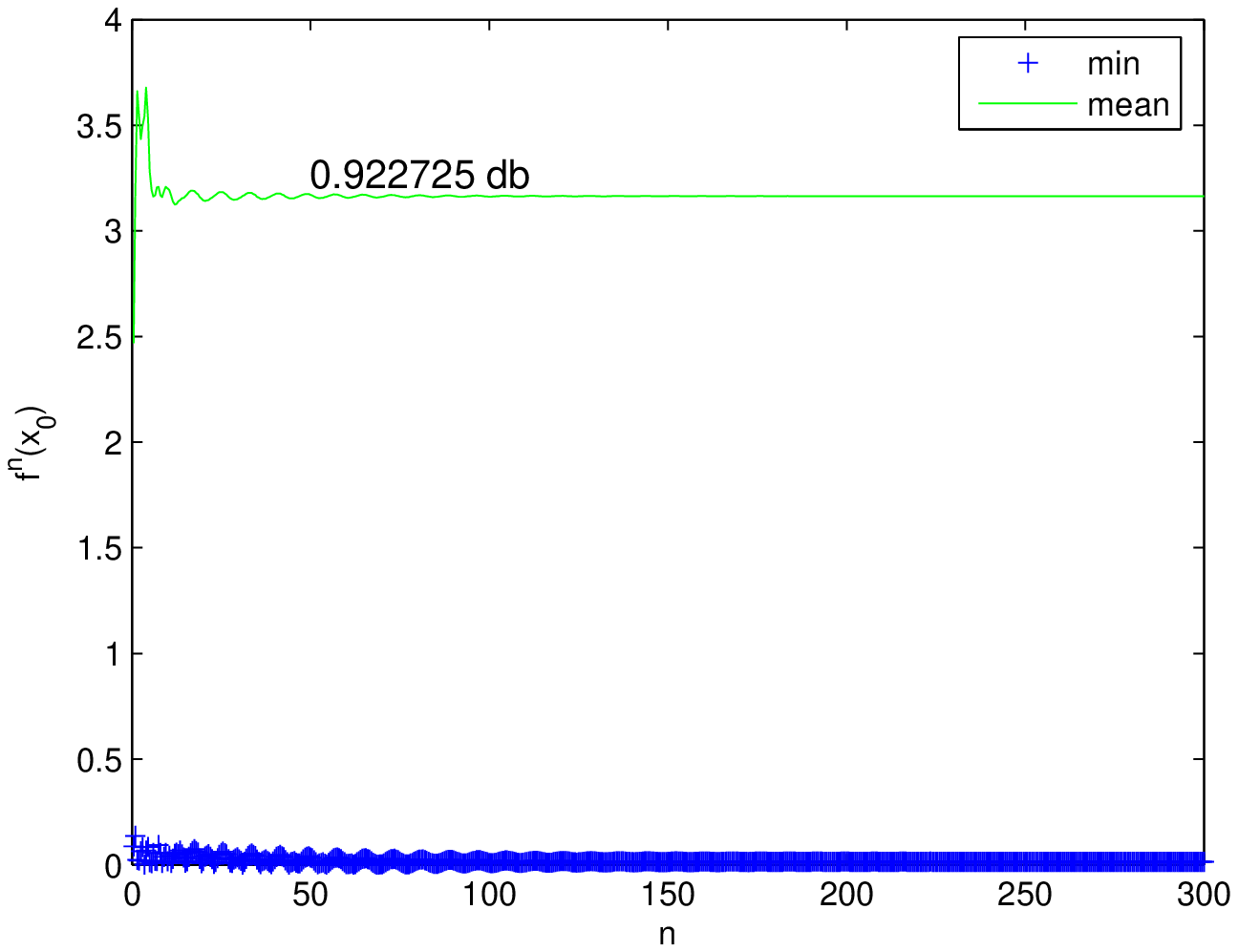} 
\includegraphics[width=1.8in]{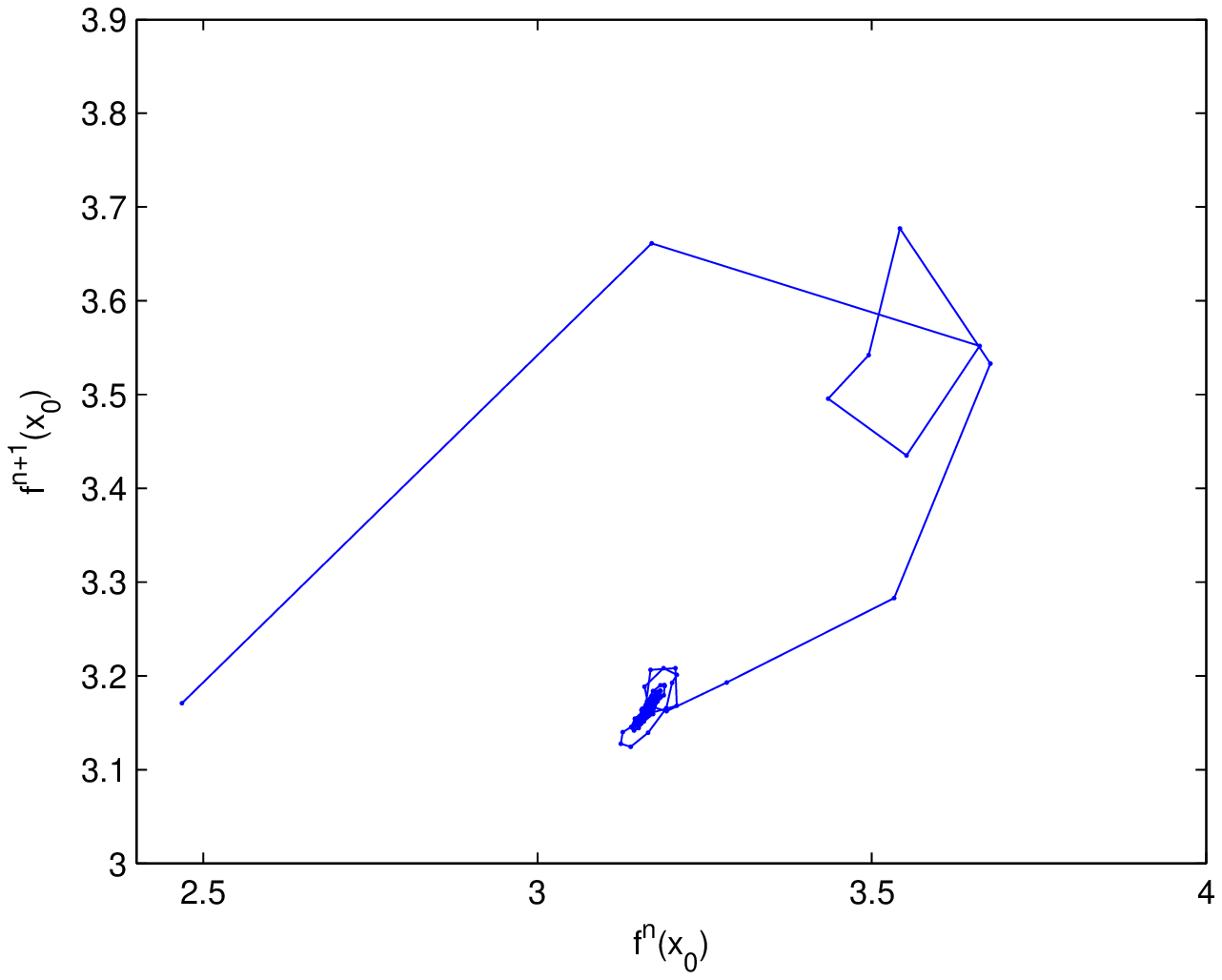}
}
\centerline{
\includegraphics[width=1.8in]{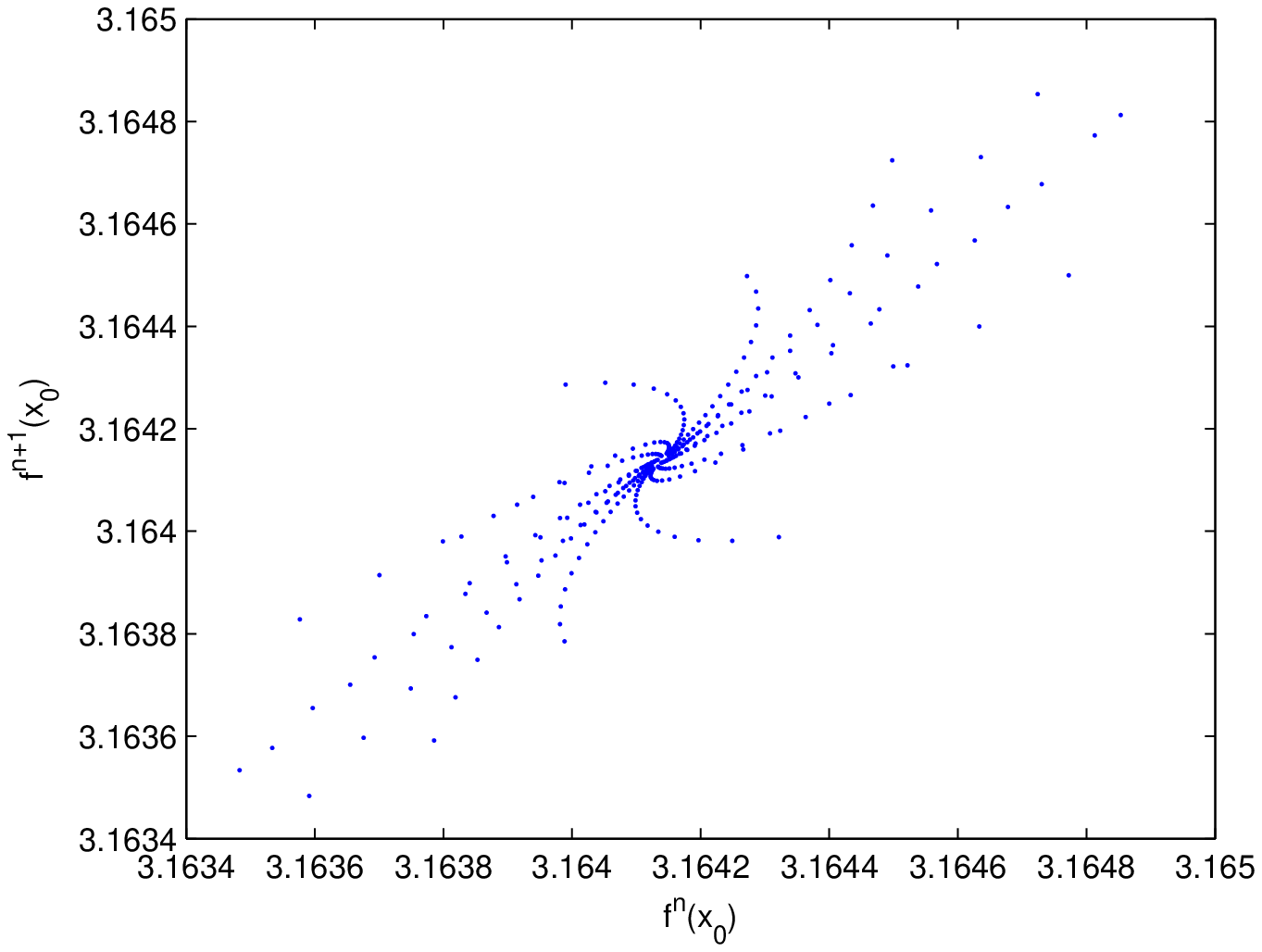}\hspace{1.8in}
}
\vspace{-0.4cm}
\caption{$\gamma=0.922725$ db, a fixed point.}
\label{fig:2f}
\vspace{0.2cm}
\centerline{
\includegraphics[width=1.8in]{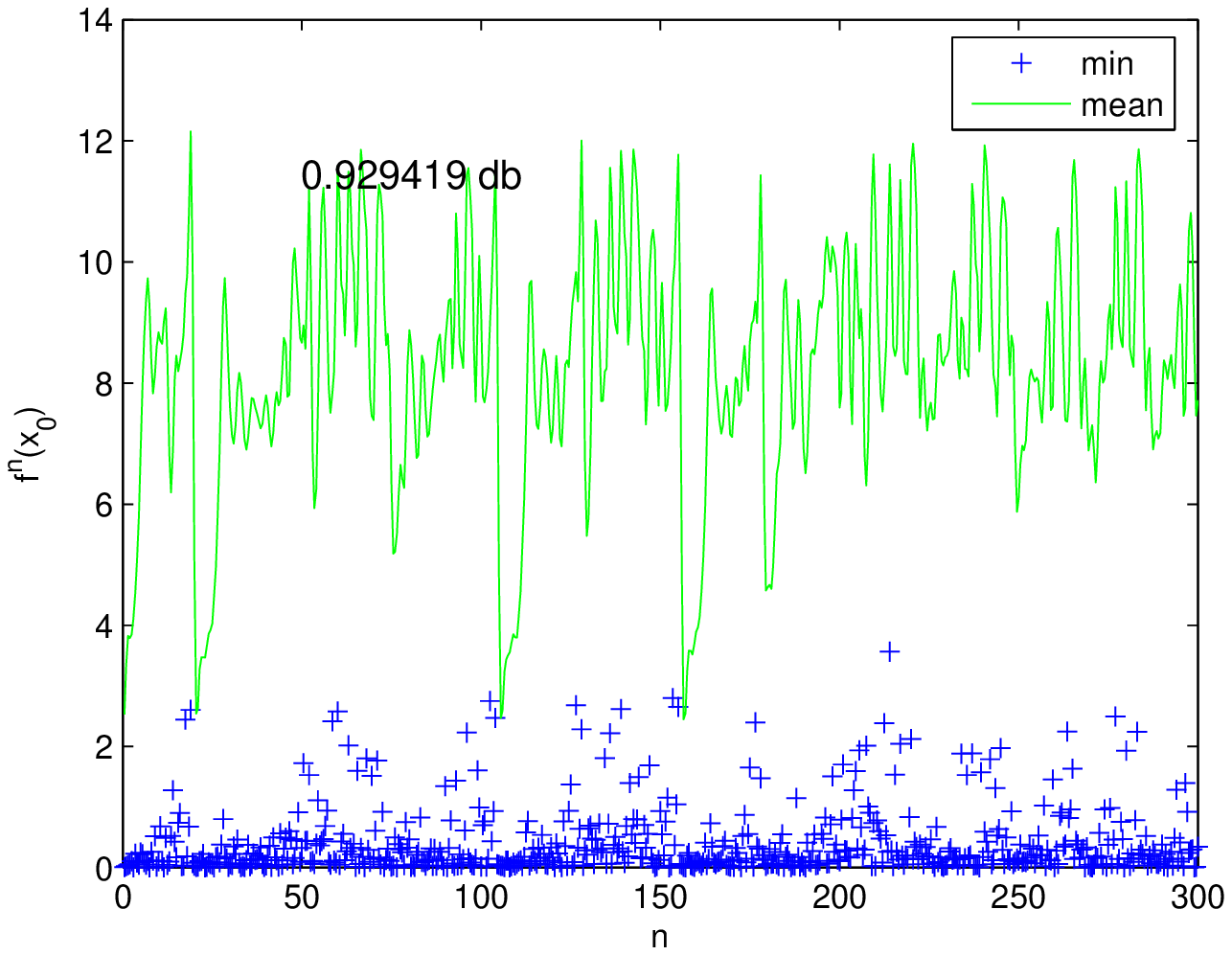}
\includegraphics[width=1.8in]{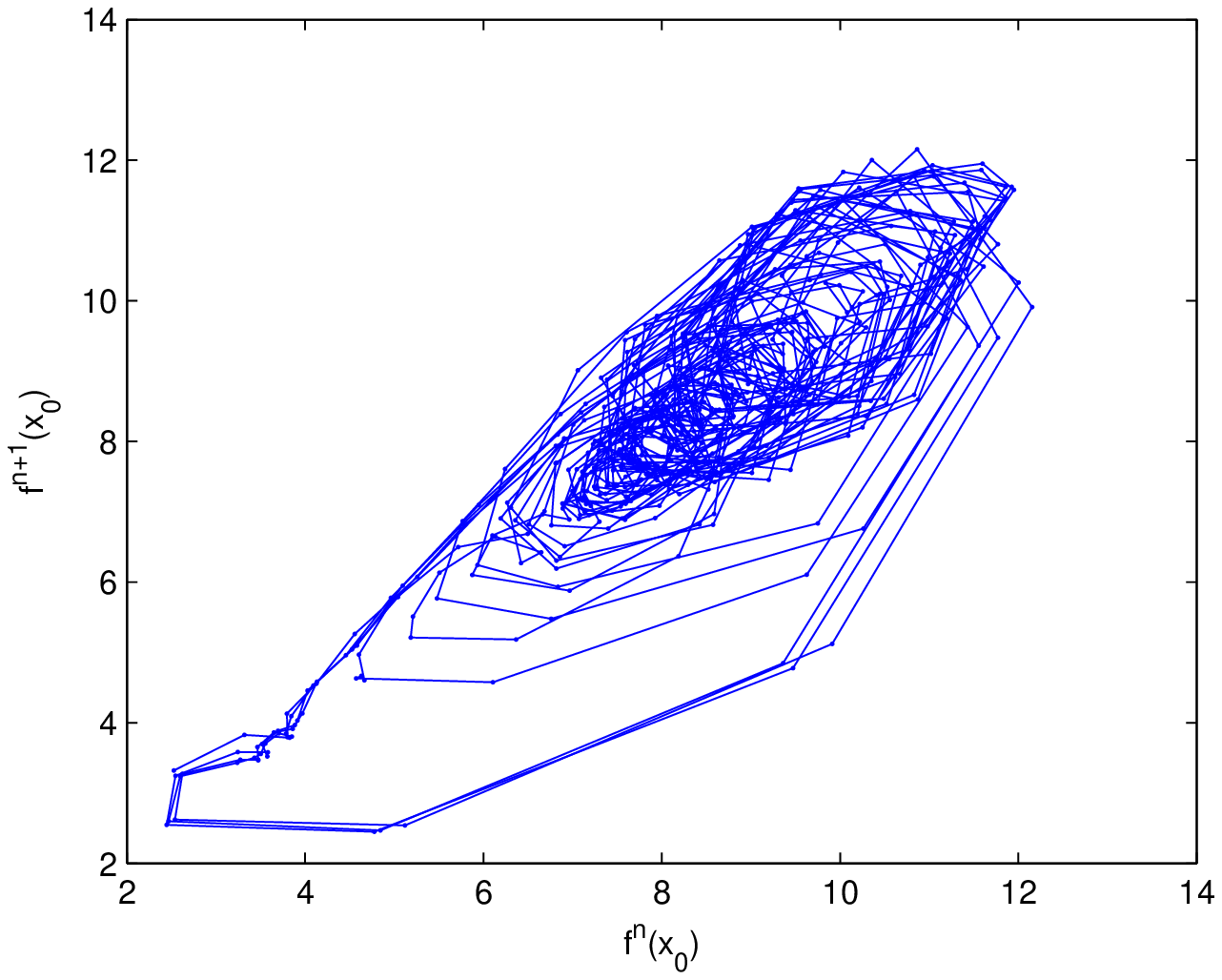}
}
\vspace{-0.4cm}
\caption{$\gamma=0.929419$ db, chaos.}
\label{fig:2g} 
\vspace{-0.4cm}
\end{figure}

\begin{figure}[htbf]
\centerline{
\includegraphics[width=1.8in]{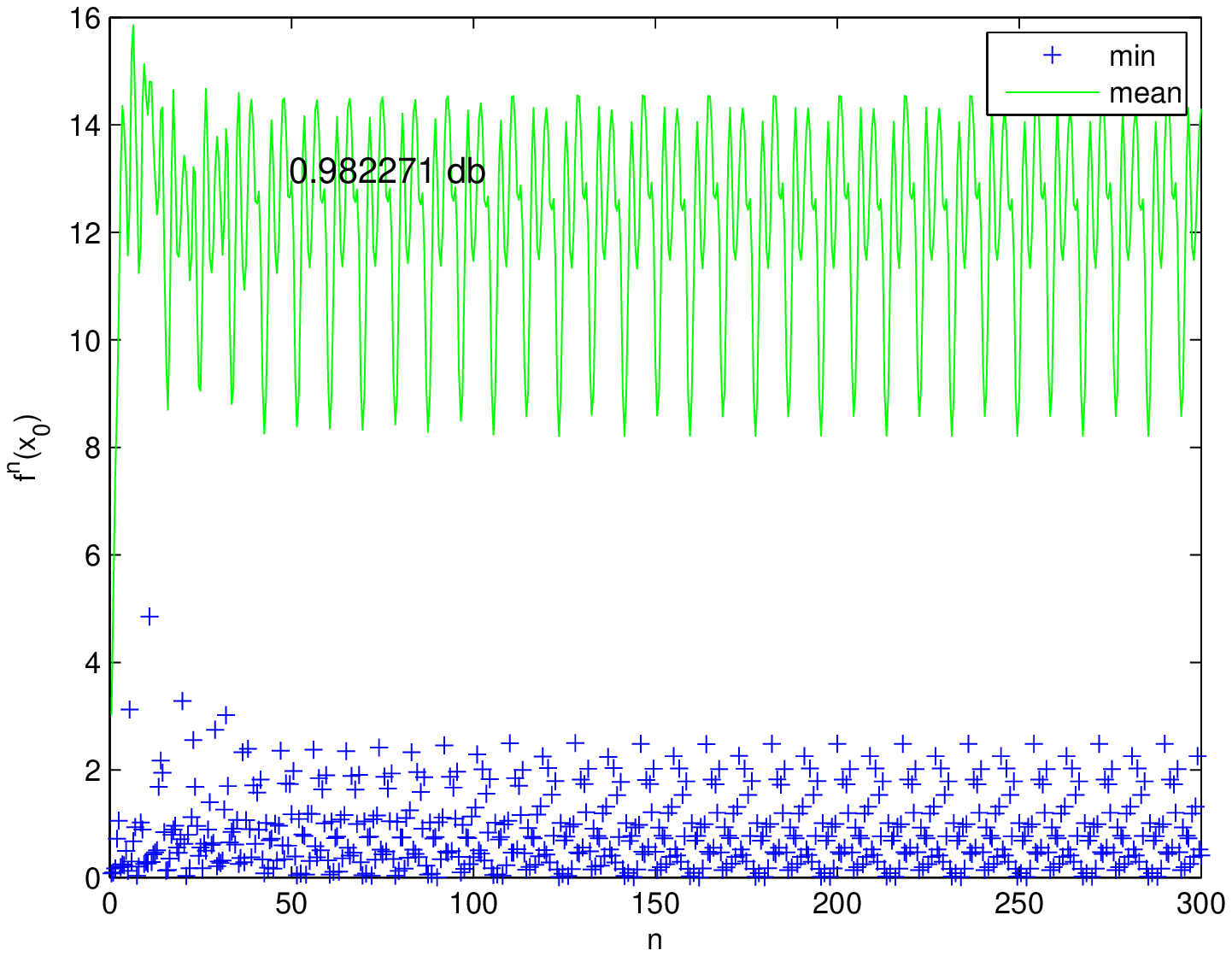}
\includegraphics[width=1.8in]{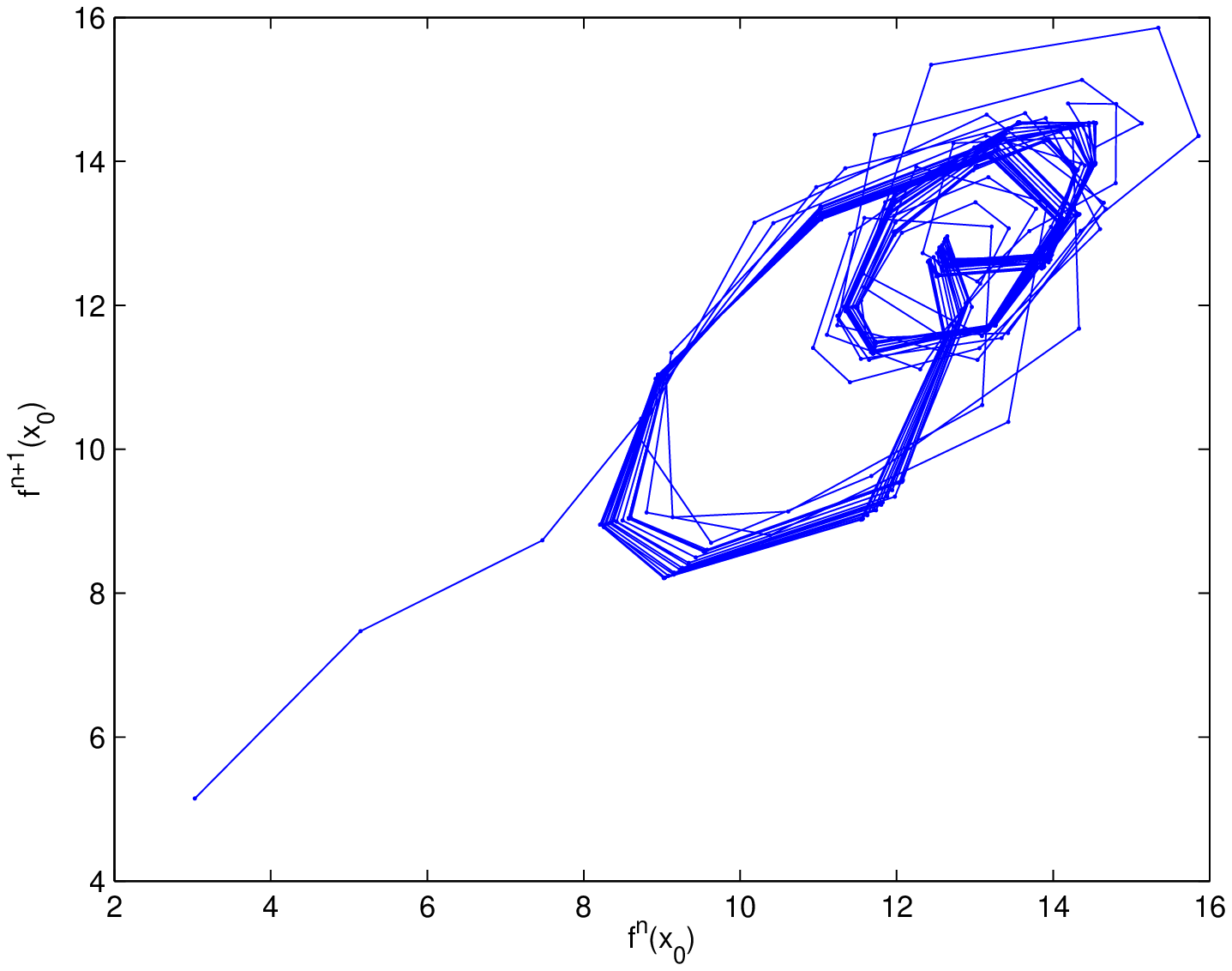}
}
\centerline{
\includegraphics[width=1.8in]{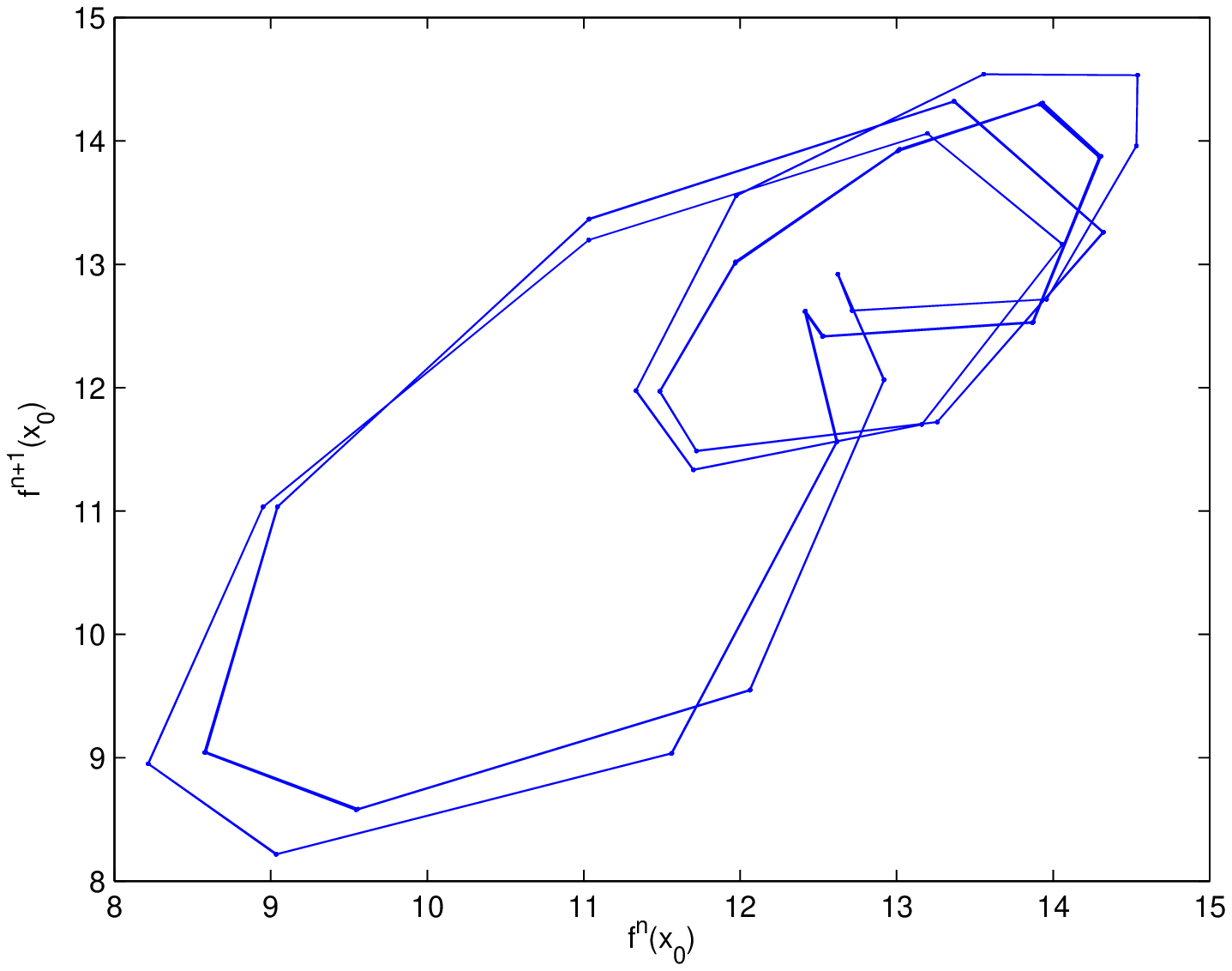}\hspace{1.8in}
}
\vspace{-0.3cm}
\caption{$\gamma=0.982271$ db, quasi-periodic fixed point.}
\label{fig:2h}
\vspace{0.2cm}
\centerline{
\includegraphics[width=1.8in]{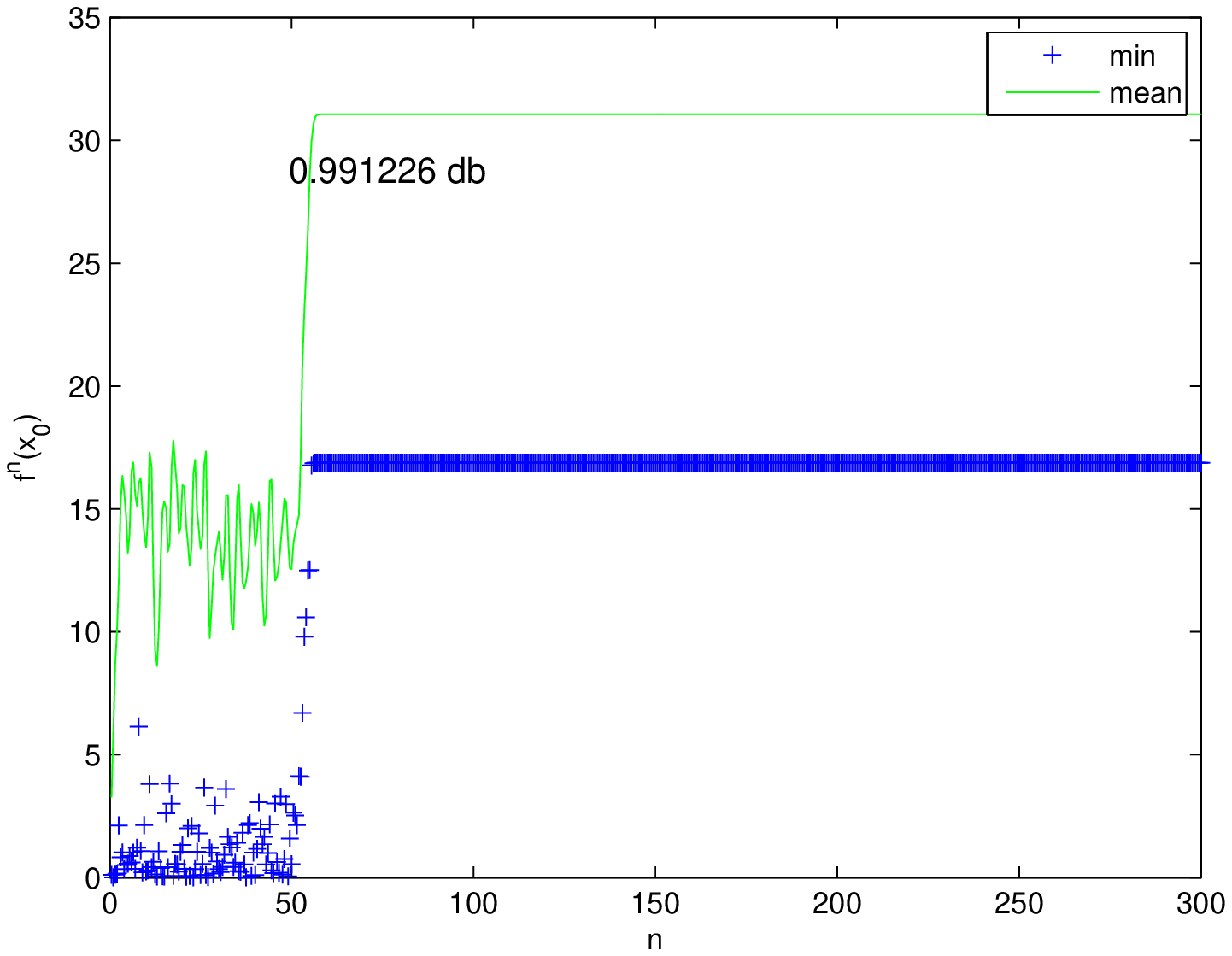}
\includegraphics[width=1.8in]{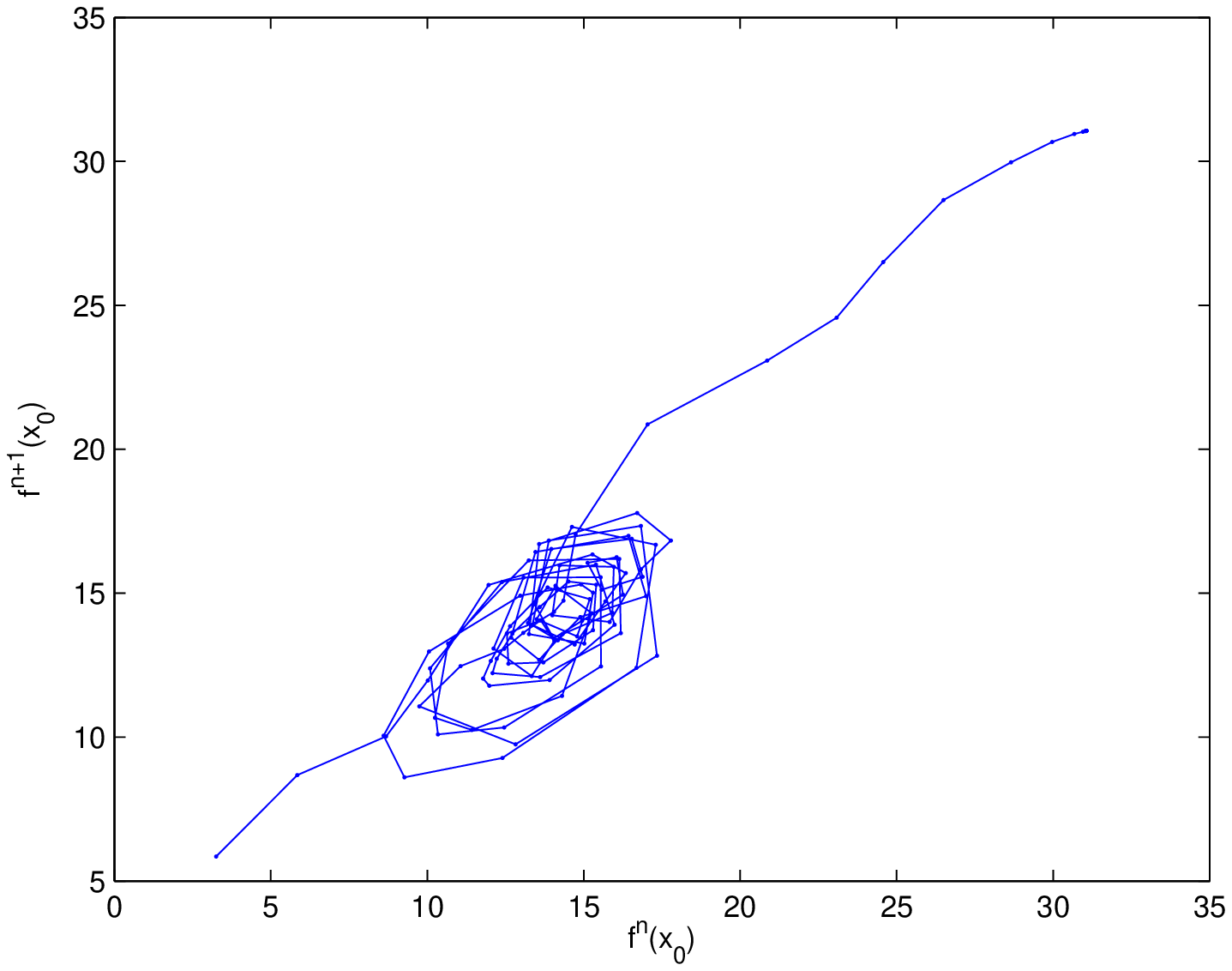}
}
\vspace{-0.3cm}
\caption{$\gamma=0.991226$ db, transient chaos}
\label{fig:2i}
\vspace{0.2cm}
\centerline{
\includegraphics[width=1.8in]{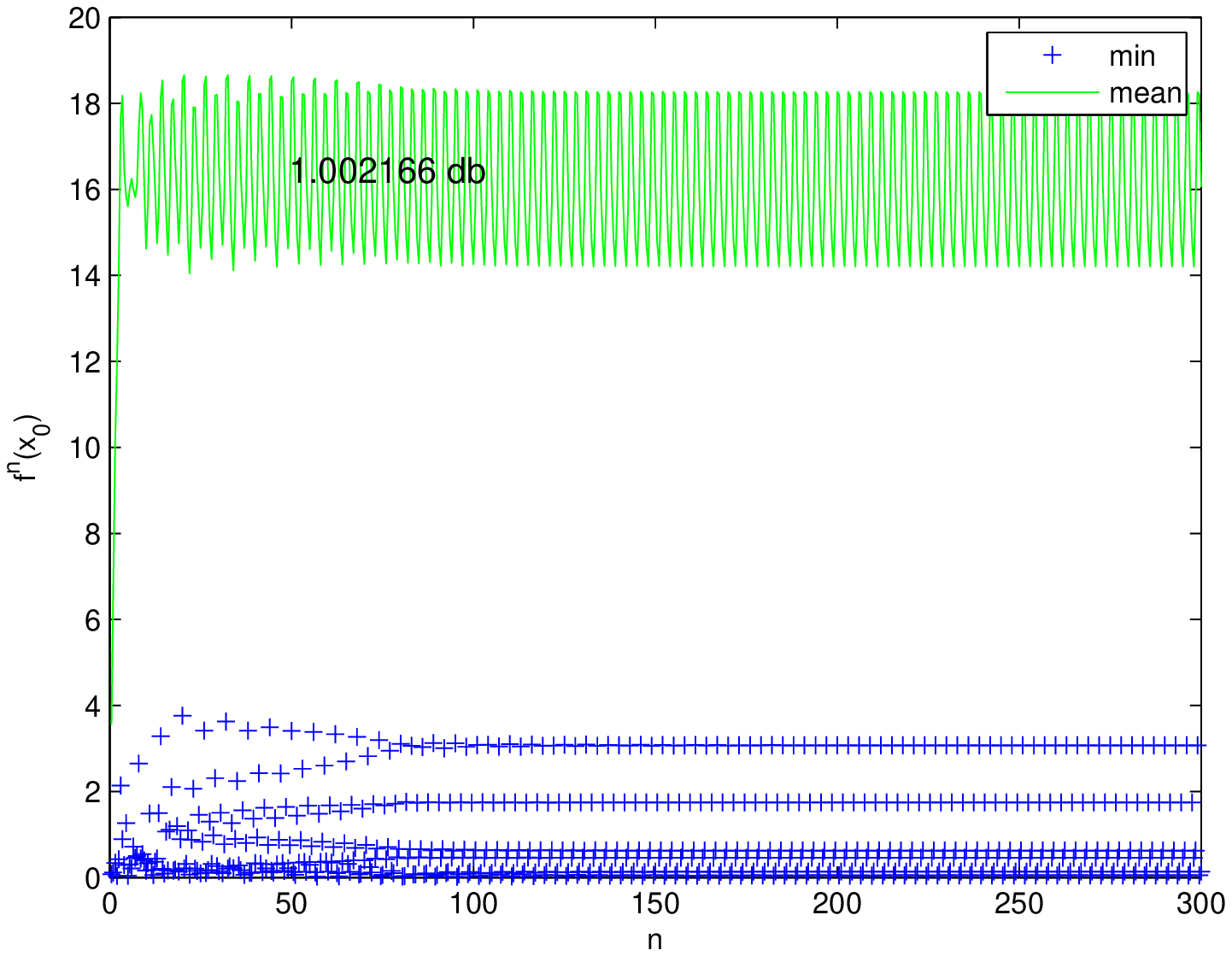}
\includegraphics[width=1.8in]{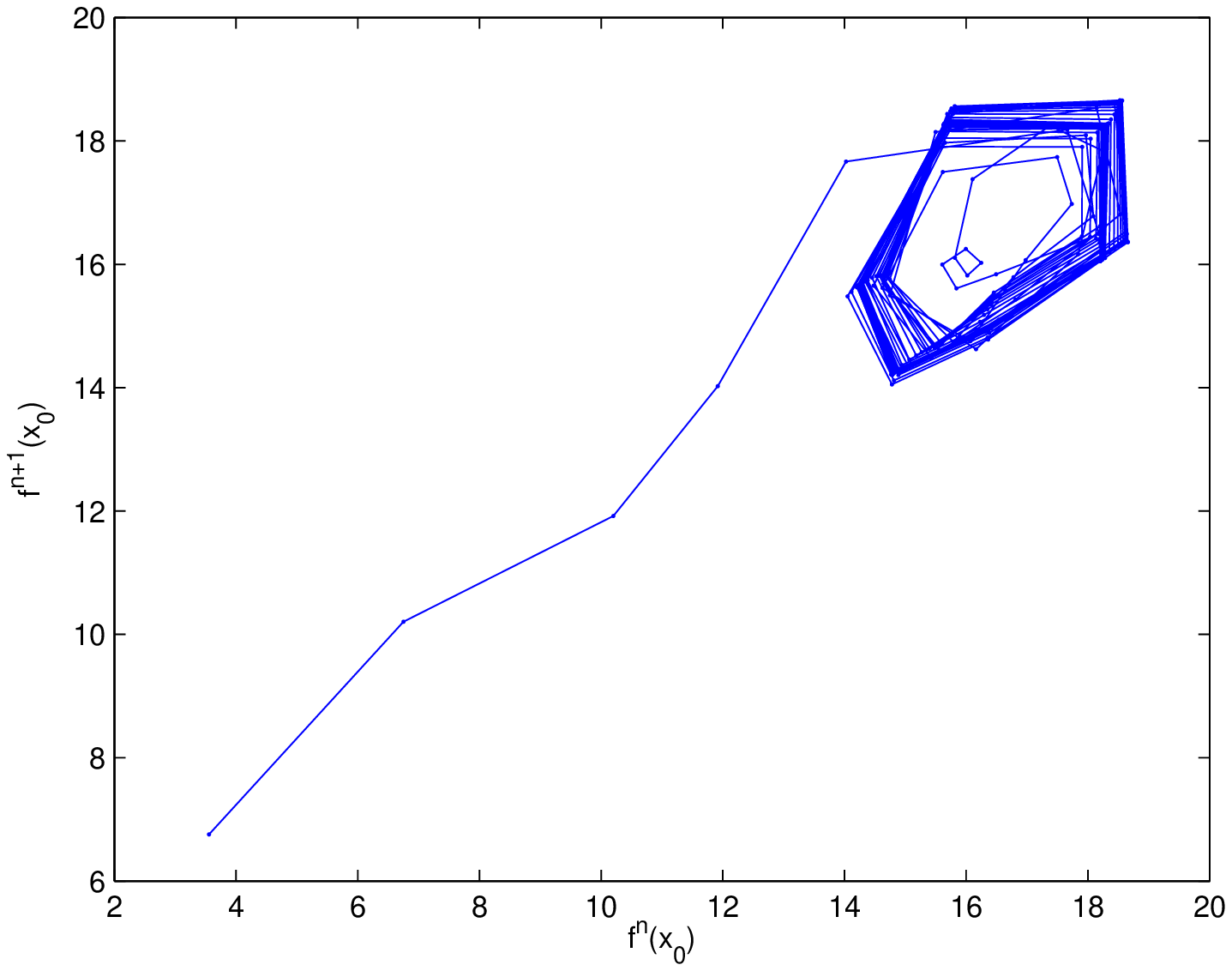}
}
\centerline{
\includegraphics[width=1.8in]{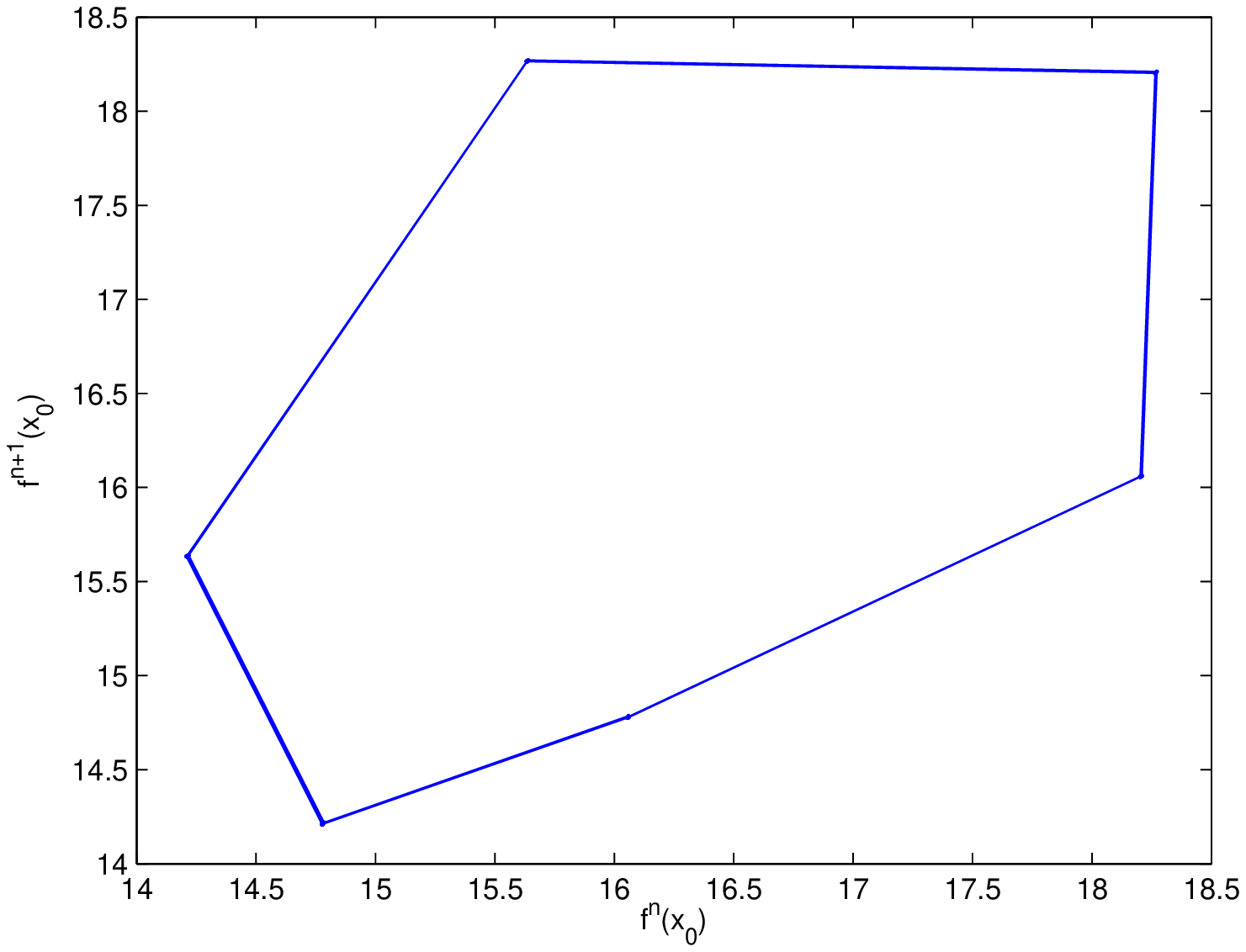}\hspace{1.8in}
}
\vspace{-0.3cm}
\caption{$\gamma=1.002166$ db, periodic fixed point.}
\label{fig:2j}
\vspace{-0.4cm}
\end{figure}

Extensive simulations are performed in our preliminary study. A whole range of phenomena known to occur in nonlinear dynamical systems, including fixed points, bifurcations, oscillatory behavior, period-doubling, limited cycles, 
chaos and transient chaos, are observed in the iterative decoding process as $\gamma$ increases (Fig. \ref{fig:2a}--\ref{fig:2i}). 

Some of these phenomena were noted in previous studies \cite{bib:agrawal1} \cite{bib:Vardy nonlinear dynamical analysis}, but we report interesting new discoveries. For each motion type, we provide two pictures: a wave picture illustrating the change of mean magnitude of LLRs $\mathbf{E(|m_u|)}$ and the minimum magnitude of LLRs  $\min(\mathbf{|m_u|})$ (y-axis) as a function of the number of half iterations (x-axis), and a  trajectory picture presenting the phase trajectory from one half iteration to the next. In some cases, we also present a third picture of a zoomed-in trajectory after 500 half iterations

\vspace{-0.2cm}
\vspace{-0.2cm}



The iterative process inevitably starts with and ends at a fixed point. The former, occurring at an asymptotically low SNR (such as $\gamma \le 0.778151$ db in our experiment,  Fig. \ref{fig:2a}), is termed an {\it indecisive fixed point}, and is associated with an unacceptably high error probability ($22\%$ in our experiment). The latter, occurring at an asymptotically high SNR (e.g., $\gamma \ge 1.113943$ db,  Fig. \ref{fig:2i}), 
 denotes a successful decoding convergence to a zero-error {\it unequivocal fixed point}.


Between the two asymptotic ends are a myriad spectrum of bifurcations, some of which correspond  well to the ostensible concepts and phenomena from the information theory, and others appear foreign and await an indepth study. As $\gamma$ increases from $0.778151$ to $0.845098$ db (Fig. \ref{fig:2b}, the system remains trapped to an indecisive fixed point, but the long convergence time 
indicates the stability of the indecisive fixed point begins to break down. 

At SNR of $\gamma=0.857332$ db, the indecisive fixed point undergoes a flip bifurcation and a stable {\it periodical fixed point} with period of 5 is formed (Fig. \ref{fig:2c}). 

Further increasing $\gamma$ to $0.902829$ db leads to an increased period from 5 to 10, showing the {\it period-doubling} phenomenon (Fig. \ref{fig:2d}). A closer inspection, shown in the zoomed-in trajectory picture, \ indicates that the motion is not exactly repetitive, but follows an approximate periodic orbit. As been verified in the experiment here (as well as other complex systems), the period will continue to double without bound as $\gamma$ increases. The ``discrete'' orbit eventually becomes continuous at $\gamma=0.921682$ db, presenting a {\it limited cycle} or {\it limited ring} -- a closed curve homeomorphic to a circle (Fig. \ref{fig:2e}). 

As SNR further increases, the limited ring loses its stability, and converges ones again to an {\it indecisive fixed point} at $\gamma=0.922725$ db (Fig. \ref{fig:2f}). This is rather surprising and is the first time that this type of indecisive fixed points has been observed for turbo decoders. Unlike the fixed points at both asymptotic ends, here  the phase trajectory oscillates with diminishing amplitude and it takes longer to converge.  

\begin{figure}[htbf]
\centerline{
\includegraphics[width=1.8in]{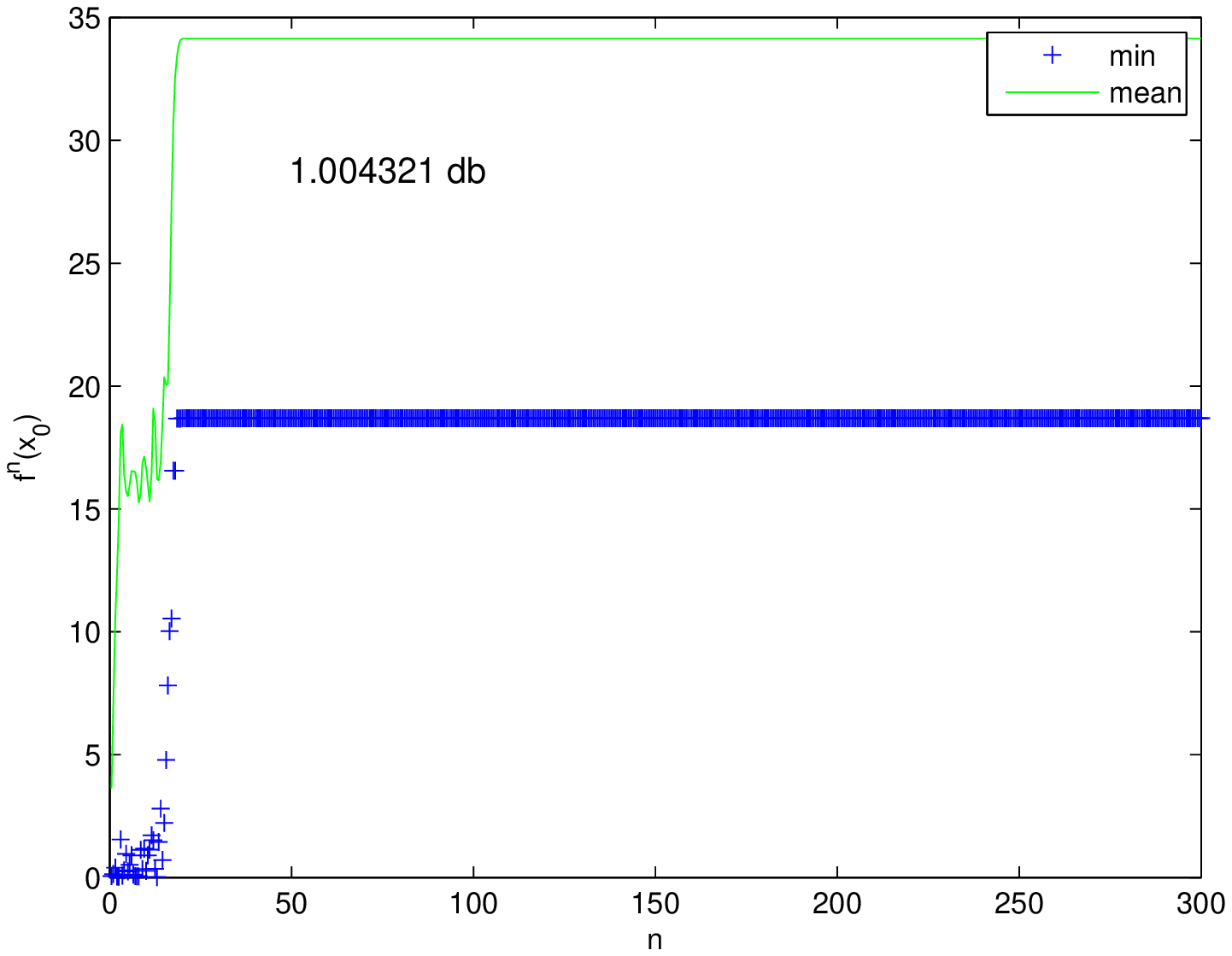}
\includegraphics[width=1.8in]{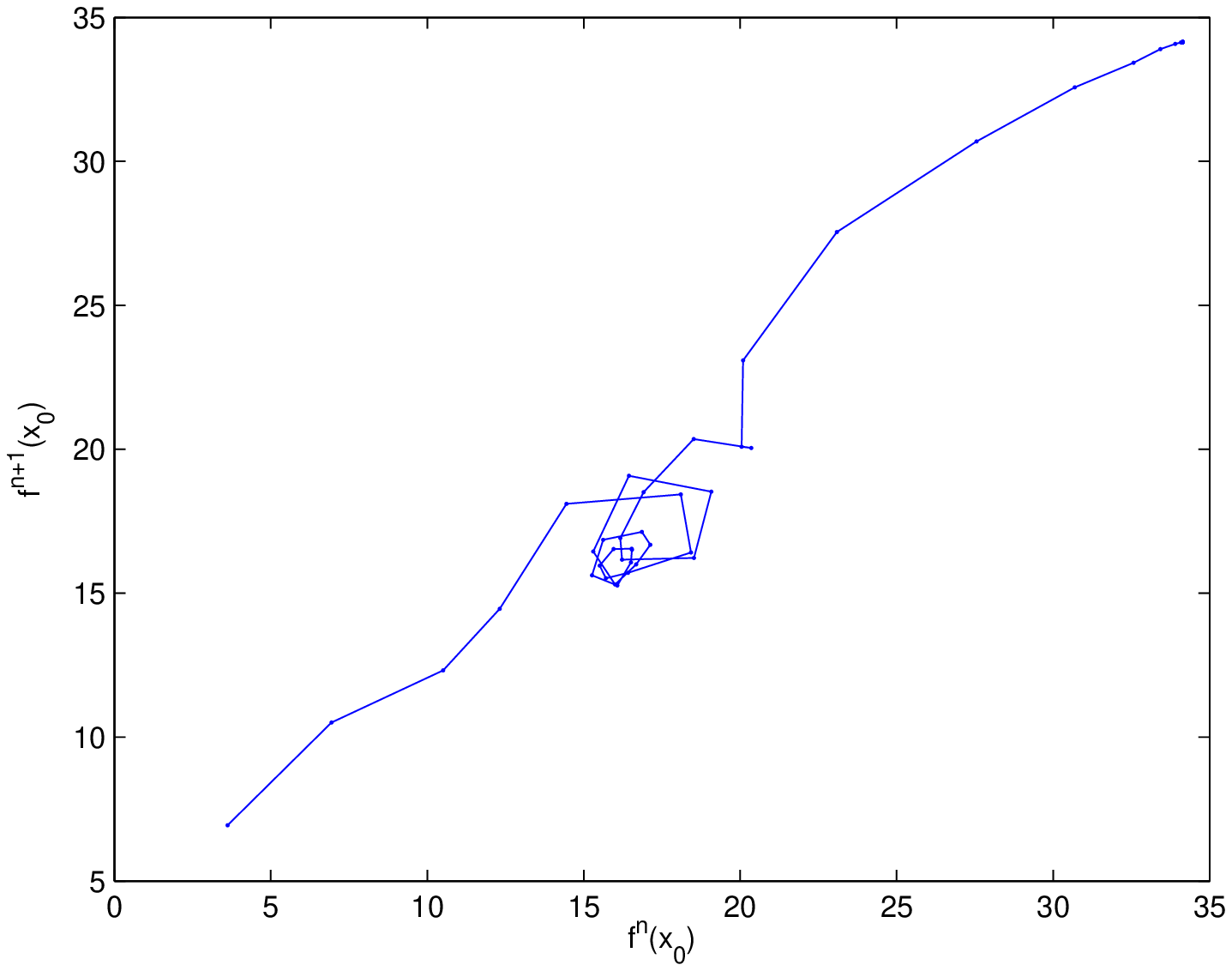}
}
\vspace{-0.3cm}
\caption{$\gamma=1.004321$ db, transient chaos}
\label{fig:2k}
\vspace{0.2cm}
\centerline{
\includegraphics[width=1.8in]{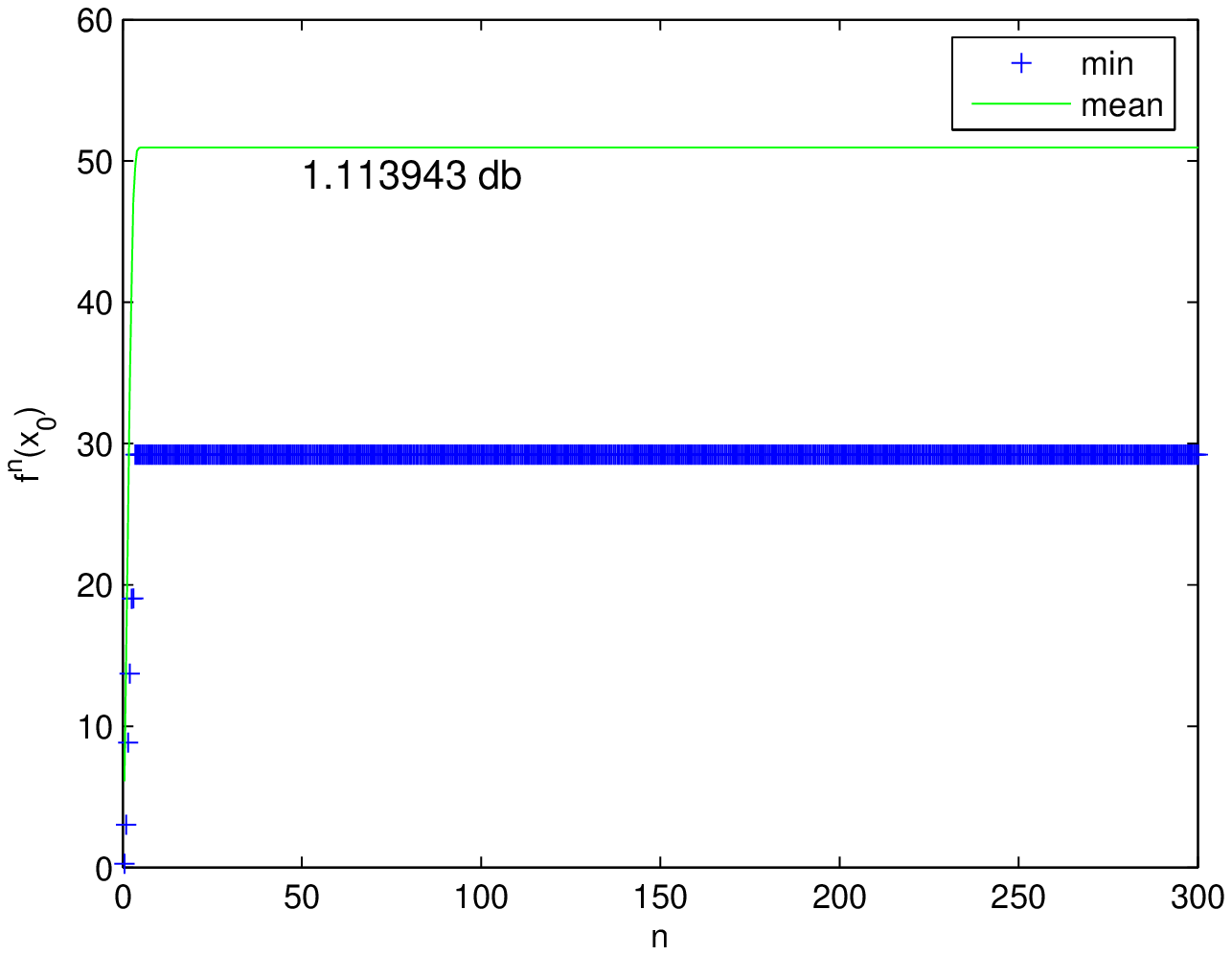}
\includegraphics[width=1.8in]{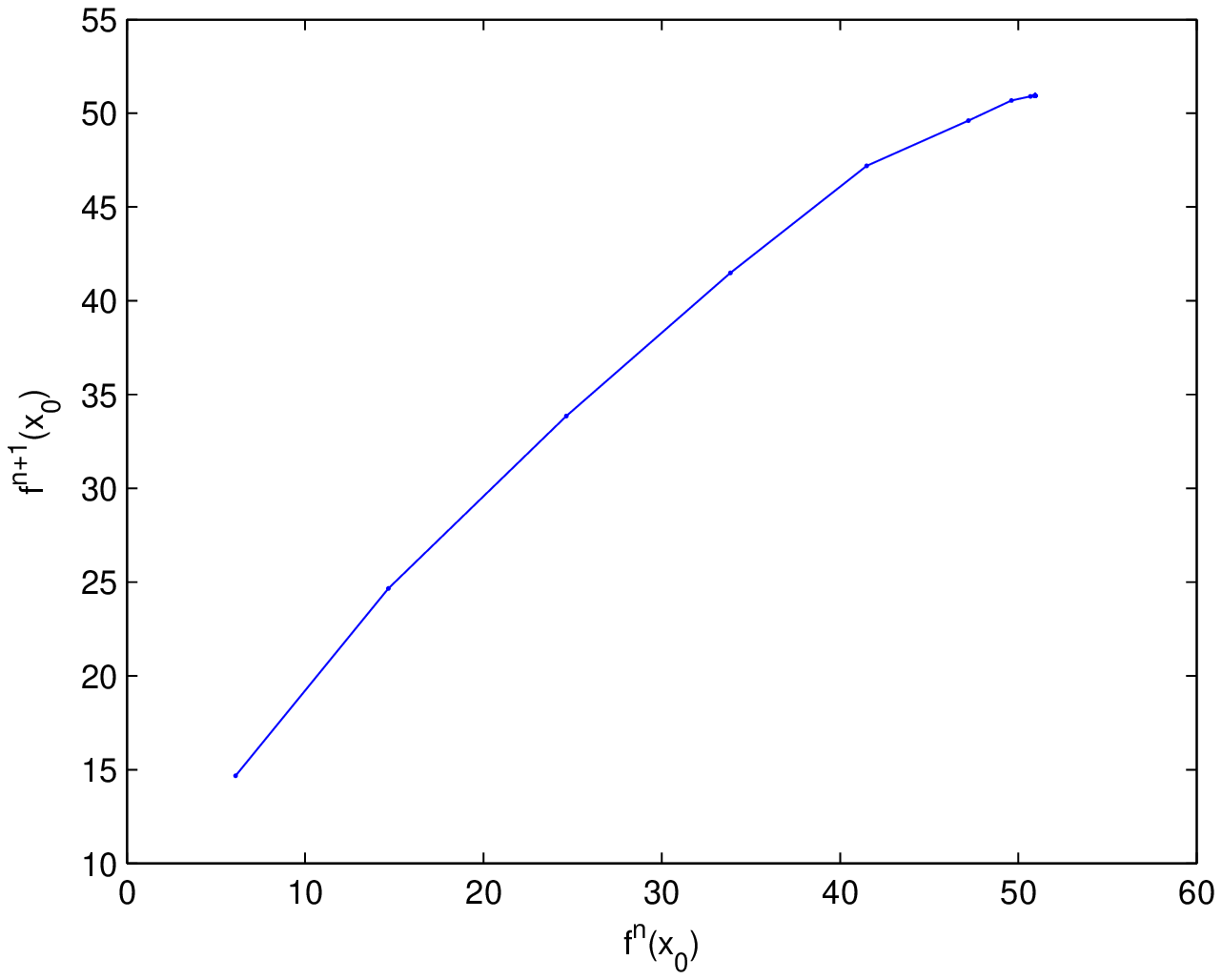}
}
\vspace{-0.3cm}
\caption{$\gamma=1.113943$ db, unequivocal fixed point.}
\label{fig:2l}
\vspace{-0.4cm}
\end{figure}

Next at $\gamma=0.929419$ db, the fixed point undergoes Neimark-Sacker bifurcation through which the phase trajectory goes into an invariant set and, after a transient period, becomes {\it chaos} (Fig. \ref{fig:2g}). 

The previous limited studies have suggested that after chaos will be transient chaos and then the convergence to an unequivocal fixed point \cite{bib:agrawal1}. It is intriguing indeed to report there actually exist a rich variety of motion types between chaos and the asymptotic unequivocal fixed point. They include: a {\it quasi-periodic fixed point} at $\gamma=0.982271$ db (Fig. \ref{fig:2h}), {\it transient chaos} at $\gamma=0.991226$ db (Fig. \ref{fig:2i}), a {\it periodic fixed point} at $\gamma=1.002166$ db (Fig. \ref{fig:2j}), another {\it transient chaos} with a short transient lifetime at $\gamma=1.00432$ db (Fig. \ref{fig:2k}), and eventually the zero-error {\it unequivocal fixed point} (Fig. \ref{fig:2l}). 

\subsection{Analysis and Simulations}

Repeated tests on a large sample of random noise realizations 
show that although different realizations
produce different bifurcation diagrams, the entire 
SNR range nonexclusively falls apart into three
regions: a low-SNR region corresponding to  
stable indecisive fixed points, a transition region known in the communication jargon as the {\it waterfall region} in which bifurcations
occur, and a high-SNR region corresponding
to stable unequivocal fixed points. It can be proven that  (1) for an iterative estimator/decoder that is probabilistic inference based, given any noise realization ${\bf z}$ and positive number $\delta$, there exists an SNR threshold $\gamma_1({\bf z},\delta)$, such that for any SNR $<\gamma_1({\bf z}, \delta)$, the iterative algorithm converges,  with a probability greater than $1-\delta$, to a unique and stable indecisive fixed point; (2) likewise,  there exists an SNR threshold $\gamma_2({\bf z},\delta)$, such that for any SNR $>\gamma_2({\bf z}, \delta)$, the iterative estimator/decoder starting with an unbiased initialization converges, with a probability greater than $1-\delta$, 
to a stable unequivocal  fixed point that corresponds to zero decoding errors. 

Also of interest is the rich variety of fixed points we observed in the waterfall region, none of which were reported previously. These fixed points behave much like  chaotic (sensitive) non-hyperbolic attractors. (Chaos is a special class of aperiodic, nonlinear dynamical phenomenon, and is characterized by a prominent feature of ``sensitivity to initial conditions.'' This feature, commonly known  as the ``butter-fly effect'', states that a small perturbation to the initial state would lead to huge and drastically different changes later on.)  
In comparison, the fixed points at the two extreme ends of SNR are  hyperbolic attractors, where distances along trajectories decrease exponentially in complementary dimensions in the ambient space. These newly observed fixed points, some or all of which may or may not occur depending sensitively on the specific noise realization, are usually associated with a few detection errors. They appear to provide support, from the dynamical system perspectives, for the information theoretic conjecture that there exists  
one or more pseudo codewords in the vicinity of a correct codeword (i.e. a few bits of Hamming distance away) \cite{bib:pseudo}. They may also correspond well to the coding concept of {\it stopping set} and {\it trapping set}, which characterize an  high-SNR undesirable convergence in the BSC (binary symmetric channel) decoding model and  the Gaussian decoding model,  respectively \cite{bib:stopping,bib:trapping}. 

It is particularly worth noting that 
 in almost all the motion stages,  the mean and the minimum magnitude of LLRs fluctuate in a rather notable manner as the number of iteration increases (see the wave pictures). Since the mean magnitude of LLR is shown to relate fairly well with the percentage of errors occurred in each frame \cite{bib:ruiyuanHu}, it is therefore reasonable to predict that the per-block error number will also fluctuate with iterations. 
For example, for a short block of 1024 bits, we have observed that the number of errors in a particularly block can easily vary between 80 and 120, in a ``quasi-periodic'' Z-crease manner. In other words, it is highly likely that an early lucky iteration may save both complexity and  $1/3$ less of erroneous bits than a longer, unlucky iteration.

Since where the decoder
stops also makes a difference in  per-block performance, questions arise as how to stop at the right iteration, and how much benefit there is. 
We propose the following rule of thumb to detect Z-crease:

i) The minimum magnitude of LLR, $\min(| {\bf m_u}|)$, is a very accurate
indicator of whether or not the iterative decoder has successfully
converged to the unequivocal attractor (i.e. the correct codeword). 
In a correct convergence (i.e. when the attractor is the
unequivocal fixed point or the unequivocal chaos attractor), the minimum magnitude and the mean magnitude of LLR will both increases with iterations. 
Otherwise, the iterative process is trapped in some local minimum (which corresponds to the indecisive fixed point, the quasi-periodical cycles, and the indecisive chaos). In such as case, the average magnitude of LLR ${\bf E}(| {\bf m_u}|)$ may continue to increase with iterations at a decent pace, but the minimum magnitude  $\min(| {\bf m_u}|)$ will remain at a very low value, sending a clear signal of unsuccessful convergence. It is thus convenient to set a threshold for  ${\bf E}(| {\bf m_u}|)$ to indicate decoding success. 

ii) The Z-crease is most prominent (with large error fluctuation) in the quasi-periodical cycle and the indecisive
chaos stages. 
Hence, it is beneficial to detect  Z-crease phenomenon as early as possible and to terminate decoding at the earliest ``best'' iteration. 
Since the Z-crease of the bit errors is almost always accompanied with a 
Z-crease of ${\bf E}(| {\bf m_u}|)$, we suggest using ${\bf E}(| {\bf m_u}|)$ to detect the Z-crease of errors. Here is a simple but rather effective method:   
 Each local maximum point of ${\bf E}(|{\bf m_u}|)$ is taken as a {\it candidate point}, which is like the ``local optimal point''.   
We predict that Z-crease is occurring  when the ${\bf E}(| {\bf m_u}|)$ value of any one candidate point is lower than the ${\bf E}(| {\bf m_u}|)$ value of its
previous candidate point. 

Following these observations, we also propose a heuristic stopping criterion 
and suggest performing iterative decoding in the follow manner: 
 The iterative decoder keeps track of $\min(| {\bf m_u}^{(n)}|)$ and ${\bf E}(| {\bf m_u}|^{(n)})$, and  terminate when any one of the
following conditions happens:
\begin{enumerate}
\item When $\min(| {\bf m_u}^{(n)}|)$ increases above a threshold.
\item When ${\bf E}(| {\bf m_u}|)$ of any one candidate point is lower than
${\bf E}(| {\bf m_u}|)$ of the previous candidate point.
\end{enumerate}

If the decoder stops at condition 1), the current bit decisions and the current iteration are considered as our ``best shots.'' If the decoder stops at condition 2), then we suggest the decoder trace back to the previous candidate point and use the bit decisions of that iteration as the final decision. 
Otherwise, the decoder will proceed to reach the maximum iteration cap without voluntary stop. 

It should be noted that previous researchers have also used the mean magnitude of LLR for early stopping purpose \cite{bib:MM}, but it was used in a different way that did not recognize the Z-crease phenomenon. To the best of our knowledge, the minimum magnitude of LLR has not been  exploited previously. Our stopping criterion here is most useful in improving the worst-case per-block performance, but not so much for the average performance (averaged over lots of blocks). It can also cut down the iteration number by $50\%$ or even larger (especially at low $\gamma$s or when the specific frame encounters lucky deep distortion), without sacrificing the average performance. 

\section{Conclusion}
\label{sec:conclusion} 

We report the Z-crease phenomenon in
soft-iterative decoding systems, and use the theory of nonlinear
dynamics to justify its existence and generality. We show that while the average system error rate performance in general improves (or, does not deteriorate) with iterations, for individual frames, more iterations may actually do harm to the decoding decisions. Analyzing the dynamical behavior of the system, we further propose a simple stopping criterion based on the minimum magnitude 
and the mean magnitude of LLR  to 
detect successful convergence and determine the right iteration to stop.


\begin{thebibliography}{99}

\bibitem{bib:DE}
S.-Y. Chung, R. Urbanke and T. J. Richardson, ``Analysis of
sum-product decoding of low-density parity-check codes using a
Gaussian approximation,'' {\it IEEE Trans. Inf. Theory}, vol.
47, pp. 657-670, Feb. 2001.

\bibitem{bib:EXIT}
A. Ashikhmin, G. Kramer, and S. ten Brink, ``Extrinsic information transfer functions: model and erasure channel properties,''   
{\it IEEE Trans. Inf. Theory},
pp. 2657-2673, Nov. 2004.







\bibitem{bib:Richardson first dynamical analysis for turbo} T. J. Richardson, ``The geometry of turbo decoding dynamics,'' {\it IEEE
Trans. Inf. Theory,} vol. 46, no. 1, pp. 9-23, Jan. 2000.

\bibitem{bib:Duan fixed points for turbo} L. Duan and B. Rimoldi, ``The iterative turbo decoding algorithm has fixed points,'' {\it IEEE Trans. Inf. Theory,} vol. 47, no. 7, pp.
2993-C2995, Nov. 2001.

\bibitem{bib:agrawal1} D. Agrawal and A. Vardy, ``The turbo decoding algorithm and its phase trajectories,'' {\it IEEE Trans. Inf. Theory,}
pp. 699-722, Feb. 2001.

\bibitem{bib:Vardy nonlinear dynamical analysis} L. Kocarev, F. Lehmann, G. M. Maggio, B. Scanavino, Z. Tasev, and A. Vardy, ``Nonlinear dynamics of iterative decoding systems: analysis and applications,'' {\it IEEE Trans. Inf. Theory},  
 pp. 166-1384, April 2006.



\bibitem{bib:He dynamic analysis for Turbo product code} 
Y. He, F. C. M. Lau and C. K. Tse, ``Study of bifurcation behavior of two-dimensional turbo product code decoders,''  {\it Chaos, Solitons and Fractals,} vol. 36, pp. 500-511, 2008

\bibitem{bib:MM}
J. Li, X-H You, and J. Li, 
``Early stopping for LDPC decoding: convergence of mean magnitude (CMM),''
{\it IEEE Commun. Letters}, pp. 667-559, Sept. 2006


\bibitem{bib:pseudo}
R. Koetter, and P. O. Vontobel ``Graph-covers and iterative decoding of finite length codes,'' {\it Proc. 3rd Intl. Symp. Turbo Codes}, 2003

\bibitem{bib:PAcodes}
J. Li, K. Narayanan, and K. Georghiades, ``Product accumulate codes: a class of codes with near-capacity performance and low decoding complexity,'' {\it IEEE Trans. Inf. Theory}, pp. 31-46, Jan 2004

\bibitem{bib:stopping}
A. Orlitsky,  K. Viswanathan, and J. Zhang, ``Stopping set distribution of LDPC ensembles,'' {\it IEEE Trans. Inf. Theory}, pp. 929-953, March 2005 


\bibitem{bib:trapping}
C. Di, D. Proietti, I. Telatar, T. Richardson, and R. Urbanke,
``Finite length analysis of low-density parity-check codes,''
{\it IEEE Trans. Inf. Theory,} pp 1570-1579, June 2002.

\bibitem{bib:ruiyuanHu}
R. Hu and J. Li, ``Exploiting Slepian-Wolf codes in wireless user cooperation,'' {\it Proc. IEEE Sig. Processing Advances in Wireless Commun. (SPAWC)}, pp. 275-279, June 2005
\end{thebibliography}
\end{document}